\documentclass[%
 reprint,
superscriptaddress,
 amsmath,amssymb,floatfix,
]{revtex4-2}

\usepackage{graphicx}
\usepackage{dcolumn}
\usepackage{comment}
\usepackage{bm}
\usepackage{color}

\usepackage[normalem]{ulem}

\usepackage{booktabs}
\usepackage{array}

\usepackage{soul}
\newcommand{\modif}[1]{#1}

\begin{document}

\title{Stick-slip dynamics in an interleaved system with self-amplified friction}

\author{A. Plati}
\author{F. Restagno}
\author{C. Poulard}
\affiliation{Universit\'e Paris-Saclay, CNRS, Laboratoire de Physique des Solides, 91405 Orsay, France}

\begin{abstract}
    Understanding how stick-slip dynamics manifests in diverse physical conditions is a crucial topic in tribology. Although it has been extensively studied in simple frictional configurations, the characterization of stick-slip behavior in complex assemblies is challenging. This work presents the first systematic investigation of stick-slip dynamics in a system with multiple contact surfaces undergoing friction amplification through conversion of traction forces into normal compression. Using interleaved paper blocks as a model system, we combine force measurements and image processing to characterize stick-slip events occurring as the two blocks are pulled apart at different detachment velocities. We find that both the peak force and the amplitude of the stick-slip events decrease along with the system's detachment. By combining a previously designed model for friction amplification and the stick-slip dynamics predicted by a simple frictional spring-block system, we link the observed behavior to the evolving normal compression within the assembly. Through force measurements and imaging, we extract the effective stiffness of the system from stick-slip events at low velocities and relate it to the system's normal compression. We then predict the observed decrease of the global stiffness as a function of the detachment by considering the spatial distribution of normal forces within the assembly, which determines an effective number of sheets contributing to the system's mechanical response.  Our findings reveal a non-trivial interplay between internal stress distribution and mechanical response mediated by frictional forces, with implications for granular materials, textiles, fibrous systems, and mechanical metamaterials.\end{abstract}

\maketitle

\section{Introduction}

Friction~\cite{Bhushan2001-wh,Vanossi2013} is present in countless aspects of daily life, allowing us to walk and experience sensory perception~\cite{shewan2020tribology}. In addition to these useful properties, friction dissipates mechanical energy as heat through potentially destructive mechanisms such as wear. This makes a fundamental understanding of friction a central topic in tribology, with important applications in industrial manufacturing processes. 
The simplest laws of friction were first formulated by Da Vinci~\cite{Hutchings2016} and then rediscovered and formalized by Amontons and Coulomb~\cite{Desplanques2015}. The modern framework of tribology was subsequently established by Bowden and Tabor~\cite{Bowden1950}.
A defining dynamical characteristic of friction is stick-slip motion~\cite{Baumberger2006}: when the coefficient of kinetic friction is lower than the static one, sliding surfaces undergo alternating cycles of adhesion and slipping with varying regularity. This nonlinear dynamics manifests across diverse phenomena in both soft and hard condensed matter systems, spanning scales from nanometric contacts to seismic faults~\cite{Baumberger1996,Scholz1998,Baumberger1999,Marone1998,BaldassarriBrownian,Rastei2013puck,dalbe2015multiscale,Marchand2020,Petrillo2020,Plati2022,Yan2023,Hu2023-mq}. Understanding how stick-slip motion manifests in diverse physical conditions (e.g., dry, wet, worn surfaces)~\cite{lu2025experimental,dong2017stick,zhang2023influence} is then crucial for a wide range of disciplines, as is the idea of controlling its dynamical features~\cite{armstrong2002stick,plati2025control}.

Standard laboratory and theoretical investigations of friction consider two surfaces in contact under applied tangential and normal forces~\cite{Rice1983,gao1994dynamic,Baumberger1996,elmer1997nonlinear,Baumberger1999}. The tangential force drives relative sliding, while the normal force, independent of the tangential one, enhances contact strength by increasing both static and dynamic friction. However, many real systems—particularly those involving granular, textile, and fibrous materials—consist of complex assemblies of deformable objects with many surfaces in contact at the same time. \modif{In these systems, applying a traction force perpendicular to the normal load can lead to a local increase in the forces perpendicular to the internal contact surfaces. This results in an effective amplification of the external normal load, which enhances the overall frictional resistance.} This geometric conversion mechanism gives rise to distinct friction amplification phenomena.
A first example is provided by the belt friction, which occurs when a deformable object such as a cable or a belt is wrapped around a rigid body like a rod, as seen in capstans used on ships. In these systems, the friction force is generated by the traction that is applied~\cite{jung2008generalized}. Recent research proposes that the capstan effect may also underlie the process by which DNA interacts with bacteriophage capsids~\cite{ghosal2012capstan}. Another illustrative case is the Chinese finger trap, which relies on a woven helical braid that tightens around the fingers as it is pulled. This braiding mechanism is relevant to surgical sutures~\cite{su2012modified} and potentially to the frictional behavior observed in F-actin filament bundles~\cite{brown2015physiological}.
Multicontact assemblies further involve complex interplay between internal stress distribution, heterogeneities, and overall frictional or cohesive behavior. Indeed, different contact surfaces in static traction or relative motion within the assembly may experience variable amounts of normal compression, depending on their specific position. For example, we can consider the distribution of internal stresses in fibrous materials, such as ropes~\cite{leech2002modelling,Allan2025-dq,Allan2025-fp} or textiles~\cite{poincloux2018geometry,poincloux2018crackling,schwartz2019structure}, when they are subjected to tensile strain.

A crucial problem in these systems is relating individual microscopic contact properties to large-scale effects, such as the onset of shear thickening or avalanches in granular systems~\cite{Duran98,Restagno2004}.
It is also important to mention that the emergent mechanical behavior related to friction amplification in multicontact assemblies has recently been shown to offer many possibilities for the design of metamaterials~\cite{franklin2012geometric,weiner2020mechanics,dreier2025beaded}.

A particularly striking example of friction amplification in a multicontact assembly is provided by interleaved books~\cite{alarcon2016self,taub2021nonlinear,taub2020assemblages}, which exhibit extreme resistance to separation when pulled apart. A simple model based on the Amontons-Coulomb laws revealed that this exceptional strength arises from the action of the pulling operator: even a minimal frictional force acting at the boundaries, is nonlinearly amplified throughout the interior of the stack~\cite{alarcon2016self,taub2021nonlinear}. 
This seemingly peculiar effect is, in fact, a generic feature that has been observed in a number of interleaved systems such as rolled ribbons~\cite{vani2025asymmetric}, fiber-reinforced granular materials~\cite{wierzchalek2025vane}, mushroom-shaped~\cite{Jiang2025-lk} or chain-granular~\cite{dumont2018emergent} interlocking assemblies and twisted yarns~\cite{seguin2022twist}. Remarkably, interleaved books have also inspired the design of tactile displays~\cite{an2024energy}, switchable clutches~\cite{luo2024switchable} and triboelectric power generators~\cite{Huang2021-qa}.

Although stick-slip motion has revealed interesting features when granular materials, textiles and elastic slab stacks yield~\cite{BaldassarriBrownian,alarcon2016self,poincloux2018crackling,Hu2023-mq,poincloux2024stick}, \modif{a systematic investigation into the relationship between stick-slip dynamics and friction amplification properties of multicontact assemblies is lacking. This article uses interleaved paper blocks as a model system to explore this relationship.} We point out that paper has proven to be a particularly reliable material for foundational studies on stick-slip~\cite{Heslot1994} \modif{and that the key role of friction in the mechanical response of book-like systems has been recently highlighted~\cite{poincloux2021bending}.} 
Moreover, earlier investigations into static friction and steady sliding in interleaved books, alongside the development of an analytical model that quantitatively matches experimental results~\cite{alarcon2016self,taub2020assemblages,taub2021nonlinear}, \modif{provide a solid foundation for the present work. While stick-slip motion was occasionally observed in these studies, it was never systematically investigated, leaving this dynamical feature largely unexplored in interleaved systems.}

The paper is organized as follows. In Sec.~\ref{sec:expset}, we describe the experimental setup and protocol. Sec.~\ref{sec:ststic} discusses the stick-slip features manifested in the force-displacement curves obtained through traction tests. Here, our results are interpreted with the aid of the mechanical model for friction amplification proposed in Ref.~\cite{alarcon2016self} and a standard stick-slip model based on a spring-block system. In Sec.~\ref{sec:defo},  we present an analysis that combines imaging and force measurements in order to understand the behavior of the global effective stiffness of the assembly. Finally, \modif{Sec.~\ref{sec:outlook} provides an overview of future perspectives, while Sec.~\ref{sec:conc} is dedicated to conclusions.} We also
provide a list of the symbols used in this paper in Table~\ref{tab:symbols}.

\section{Experimental Setup}\label{sec:expset}

\begin{figure}
\includegraphics[width=0.9\columnwidth]{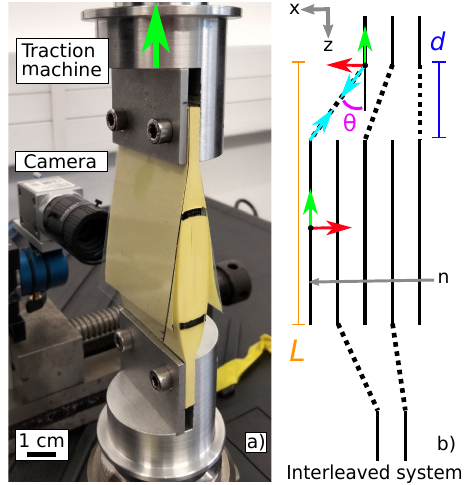}
    \caption{ a) Picture of the experimental setup. Two interleaved Post-it\textregistered~blocks are mounted on a traction machine able to pull them apart by imposing a vertical displacement. A camera is placed behind the assembly to capture images of the interleaved zone during traction tests. The black marks on the block highlight the top and the bottom interleaving points.
b) Schematic representation of the \modif{front view of half of the interleaved assembly}. An $x$-$z$ reference system is defined \modif{in this plane}. The sheets go from the clamping point to the interleaved point, forming an angle $\theta(n)$ that increases from the central to the external part of the assembly. The traction force along the z-axis (green arrow) is transmitted to the interleaving point in a direction identified by these angles (see the cyan arrows), and in turn originates an orthogonal component oriented along the negative verse of the $x-$axis (red arrow). This component then compresses the interleaved part of the assembly along the same direction. The separation distance between the clamping and the interleaved point is denoted as $d$, while $L$ is the distance between the clamping point and the end of a sheet.} 
    \label{fig:Setup}
\end{figure}

Our samples consist of couples of Post-it\textregistered~blocks (3M) carefully  interleaved sheet by
sheet on the non-glued sides. 
We cut the received squared blocks in the direction perpendicular to the glue layer, so that each one has a length of $L_t=75$ mm, width $W=45$ mm and contains $2M=50$ sheets $\epsilon=0.1$ mm thick. In our notation, we refer to single sheets on one side of the system with the index $n\in\{1,M\}$ that runs from the center of the assembly to the outer part. 
Using a press, two holes were punched in the top of the blocks so that they could be screwed into the connectors of a traction machine (MTS Criterion Model 43) equipped with a force sensor with a load capacity of 1000 N (see Fig.~\ref{fig:Setup}a). Before clamping the system, we added an outer 0.3 mm thick plastic sheet (polyethylene terephthalate) on either external side of the upper block. These sheets help to create a more reproducible configuration of the outer part of the assembly and prevent sample deterioration during repeated manipulation. To ensure greater reproducibility of the experiments, the screws were tightened using a dynamometric screwdriver with a maximum torque of 1.2 Nm. Indeed, the amount of stress imposed at the clamping point \modif{influences the initial geometrical configuration of the assembly}. Our experiments involved imposing a vertical detachment at a constant velocity $V$ on the assembly, while measuring the total traction force $F$.  The distance between the clamping point and the end of a sheet is considered constant and denoted as $L=57$ mm. As can be seen more clearly in Fig.~\ref{fig:ImageProc}a, this distance does actually vary between external and internal sheets (i.e., external sheets are more bent). \modif{However, we neglected this variation in our schematization, as it is mainly relevant for the outer pages, and its average across all the sheets is below 0.08 mm, corresponding to 0.14$\%$ of $L$.} To quantify the degree of entanglement of two blocks, we define the variable $d$, which corresponds to the separation distance from the clamping point to the interleaving point (see Fig.~\ref{fig:Setup}b). 
Upon the separation of the assembly, $d$ increases along with the total distance between the upper and lower clamping points $L+d$. Meanwhile, the interleaved length $L-d$  decreases. The traction machine allows controlling the position of the upper clamp to micrometric precision with a 50 nm resolution. In order to image the system during the detachment, a high resolution camera (Basler a2A1920 - 160umPRO, 1920 × 1200 pixels$^2$) equipped with a lens (Fujinon 12 mm 1:1.8) is placed along the direction parallel to the sheet width (see Fig.~\ref{fig:Setup}a). Once the assembly is secured in the traction machine, our experimental protocol consists of the following steps. 
First, we interleave the two blocks until we reach $d=d_0-\delta d$, where $d_0=13$ mm and $\delta d=35$ $\mu$m.  During this process, the system deforms asymmetrically because the sheets do not always slide on each other in the same way on \modif{the two horizontal sides} of the assembly. We then gently tap the two blocks \modif{on the external pages} to ensure that all the sheets fully slide on each other, making the system appear symmetric with respect to the  $z-$axis. Then, we set $d=d_0$. The system is now under tensile stress, so we gently tap again to make it relax. The force sensor is then set to zero and the measurement is started, with a vertical detachment imposed at the desired velocity $V$. This protocol ensures that different experiments start with a similar amount of residual stress in the assembly. The acquisition rate of our measurements is adjusted according to $V$ in order to reliably resolve single stick-slip events during system detachment.

During preliminary experiments, we noted a marked aging effect: for a given sample, we need to repeat the traction test up to seven times before obtaining  comparable $F(d)$ curves over consecutive runs. \modif{The measured forces are initially very high, reaching peaks of 600 N. They then stabilize at lower values, with peak forces slightly surpassing 200 N throughout the entire detachment process. This phenomenon is probably due to the aging of the contact asperities between the surfaces of the sheets, which in turn affects the paper-on-paper friction coefficient. Several studies have reported a reduction in the friction coefficient after repeated paper-on-paper friction measurements in the same verse and concerning the same contact area~\cite{borch2001handbook,de1997determination,garoff2004friction,fulleringer2014contribution}. This is exactly the condition experienced by the sheets in our interleaved assemblies during repeated traction tests. Various mechanisms have been suggested to explain this phenomenon, including structural damage to the paper surface~\cite{de1997determination}, progressive fiber orientation in the sliding direction~\cite{garoff2004friction}, and flattening of surface asperities~\cite{fulleringer2014contribution}. Beyond the effects of aging,  as paper is a hygroscopic material, humidity is also known to affect its frictional properties~\cite{borch2001handbook,kawashima2008paper}. Our experimental setup does not allow for humidity control, however, we measured  ambient  relative humidity using a commercial indoor digital thermo-hygrometer placed close to the traction machine before each measurement.
In the following, all the data analyzed in  Sec.~\ref{sec:ststic} have been acquired under similar aging histories (between 10 and 13 preliminary traction tests in the range $d\in[13,18]$ mm) and  with an ambient relative  humidity of 40 $\%$. Reproducibility tests for this dataset were conducted under similar humidity conditions, confirming the reliability of the experimental protocol.  In Sec.~\ref{sec:defo}, we also compare data obtained from different aging histories and relative humidity conditions. As for temperature, all our experiments have been performed between 20 and 23 degrees Celsius. Finally, to assess the potential contribution of electrostatic charges to the adhesion between the sheets, we performed dedicated preliminary tests using an electrostatic discharge gun to neutralize surface charges before assembling the blocks. No significant difference was observed in the resulting force-displacement curves, suggesting that electrostatic effects are negligible under our experimental conditions.}

\section{Stick-slip dynamics}\label{sec:ststic}

\begin{figure*}
\includegraphics[width=0.9\textwidth]{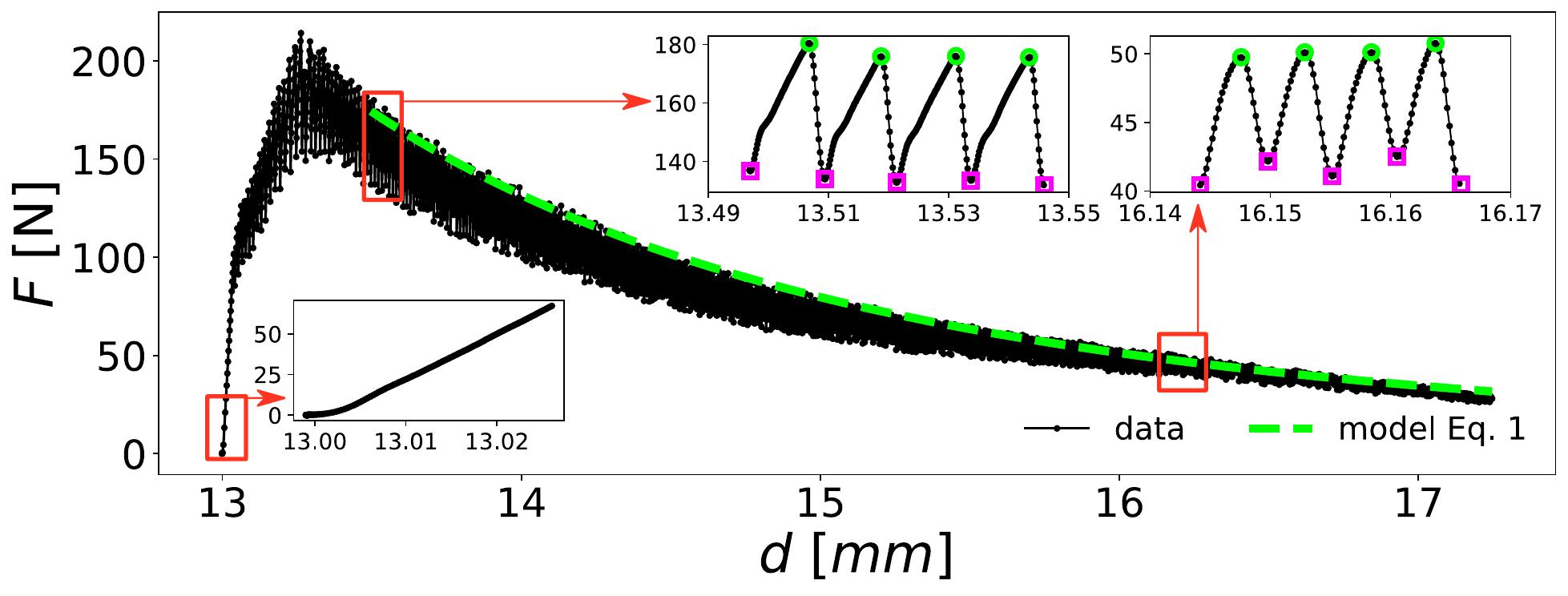}
    \centering
    \caption{Measured total traction force as a function of the separation
    distance for an experiment performed with pulling velocity $V=10$ mm/min. From left to right, the insets show the system's initial elastic response, typical early-stage stick-slip events (high amplitude, asymmetric) and typical late-stage stick-slip events (low amplitude, symmetric). Green circles refer to local maxima $F_{\text{max}}$ within single stick-slip events (also defined as $F_p=F(d_p)$ in Sec.~\ref{sec:ststic:stFeat}), pink squares refer to local minima defined as $F_v=F(d_v)$ in Sec.~\ref{sec:ststic:stFeat}. The green dashed line in the main panel shows the result of fitting $F_{\text{max}}(d)$ via Eq.~\eqref{eq::model} with $T^*=3$ mN and $\mu=0.89$.}
    \label{fig:StickSlip}
\end{figure*}

\subsection{Force-displacement curve}
Our main dataset consists of traction tests in a detachment range of $d\in[13,18]$ mm, for four values of $V=\{0.5,1,5,10\}$ mm/min.  Here we start by discussing the main features of a typical force-displacement curve $F(d)$ obtained for a given velocity (see Fig.~\ref{fig:StickSlip}). 
The traction force initially increases, exhibiting a linear behavior (see bottom-left inset). After a few tens of micrometers, the system starts to exhibit stick-slip behavior, with an average increase in force until a maximum is reached. Through various experiments, we found that the occurrence of stick-slip in the initial stage of the traction test, during which the system accumulates \modif{elastic potential energy}, depends on sample preparation. We interpret this regime as the system globally accumulating \modif{energy}, but with some local release due to residual heterogeneities in the initial configuration of the assembly. After the peak force, the system begins to detach progressively, still exhibiting stick-slip events, albeit with decreasing force on average. Resolving individual stick-slip events reveals that, contrary to the response observed in other complex frictional assemblies~\cite{poincloux2018crackling}, the overall stick-slip motion appears strikingly regular after a few events exhibiting fluctuating amplitudes. 
The two upper-right panels of Fig.~\ref{fig:StickSlip} show a zoom-in of single stick-slip events at early and later stages of detachment. Here we observe an evolution of the stick-slip features: in the first part of the detachment, the stick-slip amplitude is relatively large, and the force-displacement curve within a single event is asymmetric. In contrast, at larger $d$, the stick-slip amplitude is reduced, and the curve is more symmetric. By following the system dynamics with the camera (see Sec.~\ref{sec:video}, and the Video in the Supplementary Material) we checked that, during the stick phase, the system deforms without changing its interleaved length $L-d$ (i.e. none of the sheets are sliding), while during the slip all the pages slide at the same time (within the resolution of our measurements) thus reducing $L-d$. In other words, as in standard frictional model (i.e. spring-block systems), \modif{potential energy} is accumulated upon system deformation and subsequently released by sliding.

\subsection{Mechanical model for friction amplification}\label{sec:ststic:model}

Contrary to standard scenarios where stick-slip dynamics is stationary in time, here, after each slip event, the system results to be less interleaved than before. As reported in previous studies, the frictional properties of the assembly are crucially dependent on how much the two blocks are entangled~\cite{alarcon2016self,taub2021nonlinear}. In particular, a model was proposed to estimate the total frictional force $\mathcal{T}$ that the system opposes to detachment:

\begin{equation}\label{eq::model}
    \mathcal{T}=2MT^*\sqrt{\frac{\pi}{4\alpha}}\exp(\alpha)\text{erf}(\sqrt{\alpha}),
\end{equation}
where $\alpha=2\epsilon\mu M^2/d$, $\mu$ is the \modif{static} friction coefficient between the sheet surfaces and $T^*$ is the typical friction force due to adhesion between two pages. \modif{In the absence of stick-slip and with steady sliding assumed between all the sheets, Eq.~1 can also be used considering $\mu$ as a dynamic friction coefficient~\cite{alarcon2016self}.  However, we will never use this model based on the assumption of steady sliding.}  \modif{Eq.~\eqref{eq::model} tells us that the total traction force results from the amplification of $T^*$ (which is expected to be small) by a large factor that scales with the exponential of the square of the number of pages. For large separation distances $d \gg 2\epsilon\mu M^2$ we enter the limit where $\alpha \to 0$ and $\mathcal{T}\to2MT^*$, corresponding to the reference non-amplified case of $2M$ independent flat sheets with local friction $T^*$. This model was derived assuming standard dry Coulomb friction between paper sheets, thus neglecting rate dependence and contact aging. Nevertheless, it has been successfully  validated against experimental data in Refs.~\cite{alarcon2016self, taub2021nonlinear} by varying the number and dimensions of the sheets under an applied compression \modif{perpendicular to the traction force}. These previous studies also clarified that the friction force $T^*$ (typically of the order of mN) arises from the adhesion forces between the pages.} \modif{Any electrostatic effect on friction amplification, if present, enters the model as an adhesion contribution, which is then subsumed into $T^*$.}
We point out that Eq.~\eqref{eq::model} was derived based on the fact that the system's geometry allows for a partial conversion of the vertical force applied by the traction machine into a \modif{perpendicular} component $N$ acting on the interleaved part of the assembly (see the sketch in Fig.~\ref{fig:Setup}b). This conversion occurs for every sheet, each one feeling the contribution of all the outer sheets, resulting in an increasing local $N$ going from outside to the internal part of the assembly. 
The model predicts a normal force felt by the sheet $n$ due to the outer ones which reads:
\begin{equation}\label{eq::normal}
    N(n)=(T^*/2\mu)[\exp{[\alpha(1-(n/M)^2)]}-1].
\end{equation}
The above equation is obtained as a direct extension of the calculations performed in Ref.~\cite{alarcon2016self}, which made use of the continuum limit $1/M\ll 1$. From Eq.~\eqref{eq::normal}, we can also estimate the average normal force felt by the assembly as a whole $\mathcal{N
}=4M\int_0^1dsN(Ms)\sim \mathcal{T}/\mu$, where the integration variable $s=n/M$ is used to perform the continuum limit.

This model was originally used in studies that focused on verifying the predicted scaling with $\alpha$, rather than on resolving the stick-slip dynamics, as in our case. 
The model does not consider the presence  of stick-slip, however, Eq.~\eqref{eq::model} predicts the force needed to overcome the total static friction (considering all the sheets in static contact) given a certain geometric configuration of the assembly (which is parametrized by $d$). Within standard stick-slip models~\cite{elmer1997nonlinear}, the external traction force is equal to static friction at the start of the slip. This suggests using Eq.~\eqref{eq::model} to describe the evolution of the peak forces $F_\text{max}$ of stick-slip events as a function of $d$. In Fig.~\ref{fig:StickSlip}, we show the good agreement between Eq.~\eqref{eq::model} and the measured $F_\text{max}(d)$. This agreement is obtained by using $\mu=0.89$ and $T^*=3$ mN, which is of the same order as the one obtained in Ref.~\cite{taub2021nonlinear}. Provided a rescaling with the sheet width, our $T^*$ is also compatible with the estimate based on paper-on-paper work of adhesion provided in the same reference. On the other hand, the fitted value of $\mu$ cannot be simply traced back to a single representative paper-on-paper static friction coefficient. Indeed, as already noted in previous studies~\cite{crassous1999humidity,alarcon2016self}, adhesion forces introduce a divergence of the static friction coefficient for low normal forces between two surfaces (corresponding to a plateau of the friction force at zero normal force). 
Since we expect a non-negligible variation in the normal force between external and internal sheets according to Eq. ~\eqref{eq::normal}, we decided to quantify the relevance of this effect for our system.  In Appendix~\ref{app:singlesh}, we show the measured paper-on-paper static friction coefficient for a couple of sheets in contact under a normal load varied between 0.1 and 2 N, obtaining static friction coefficients between 1.2 and 0.54.  
The friction coefficient $\mu=0.89$ obtained by fitting $F_\text{max}(d)$ via Eq.~\eqref{eq::model} falls within the range of these values. \modif{A more comprehensive discussion on this comparison is provided in Appendix~\ref{app:singlesh}.} Furthermore, the fitted $\mu=0.89$ is close to the one obtained in Ref.~\cite{alarcon2016self} from experiments involving interleaved blocks of different dimensions and numbers of pages.
The good agreement between experimental data and Eq.~\eqref{eq::model} shown in Fig.~\ref{fig:StickSlip} demonstrates 
that the evolution of the maximum force achieved in a single stick-slip event can be explained by static properties alone, without considering the system's deformation.

\subsection{Analysis of stick-slip features} \label{sec:ststic:stFeat}
We now focus on features of stick-slip events that, contrary to $F_\text{max}$, cannot be directly interpreted based on the model developed in Ref.~\cite{alarcon2016self}.
\subsubsection{Stick-slip in a simple spring-block model}
 Before analyzing the experimental data, it is useful to recall the phenomenology of the simplest possible model exhibiting stick-slip motion namely a block of mass $m$ under a normal load $w$ and dragged through a spring of stiffness $k$  on a surface characterized by a static friction coefficient $\mu_s$ and a dynamic one $\mu_d<\mu_s$. The free end of the spring is assumed to move at a fixed velocity $v$ in the direction tangential to the surface.  
The elongation $u(t)$ of the spring with respect to its repose length is the only degree of freedom of the system, which also defines the elastic force $f(t)=ku(t)$  transmitted by the spring. 
This model exhibits stationary periodic stick-slip motion~\cite{elmer1997nonlinear}: in the stick phase, the block is completely at rest and $f(t)$ increases linearly with time until reaching $f(t)=f_{\text{max}}=\mu_s w$; then, the block slips moving under the combined effect of dynamic coulomb friction $\text{sign}(v)\mu_d w$ and elastic force until it stops. This marks the start of the next stick phase. In the low velocity limit $v\ll (\mu_s-\mu_d)w/\sqrt{mk}$, we have a simple analytical prediction for all the relevant stick-slip features, namely the stick time $t_{\text{st}}=2(\mu_s-\mu_d)w/kv$, the slip time $t_{\text{sl}} =\pi \sqrt{m/k}\ll t_{\text{st}}$ and the spring elongation at the end of the slip $u_0=(2\mu_d-\mu_s)w/k$. From these quantities we can easily obtain also the force amplitude of stick-slip events $\Delta f=\mu_s w-ku_0=2(\mu_s-\mu_d)w$. Already from these simple considerations, it is clear that three key control parameters to characterize stick-slip motion are $k$, $w$ and $v$. They are also relevant for more elaborate mechanical models of stick-slip, including, for example, a rate- and state-dependent friction coefficient~\cite{Rice1983, gao1993fundamentals}. For this reason, spring stiffness, normal load and pulling velocity are the main quantities that are varied in standard spring-block experimental settings~\cite{Heslot1994,Baumberger1999}.

\subsubsection{Force-displacement curves within stick-slip events}\label{sec:ststic:stFeat:2}
\begin{figure}
\includegraphics[width=0.99\columnwidth]{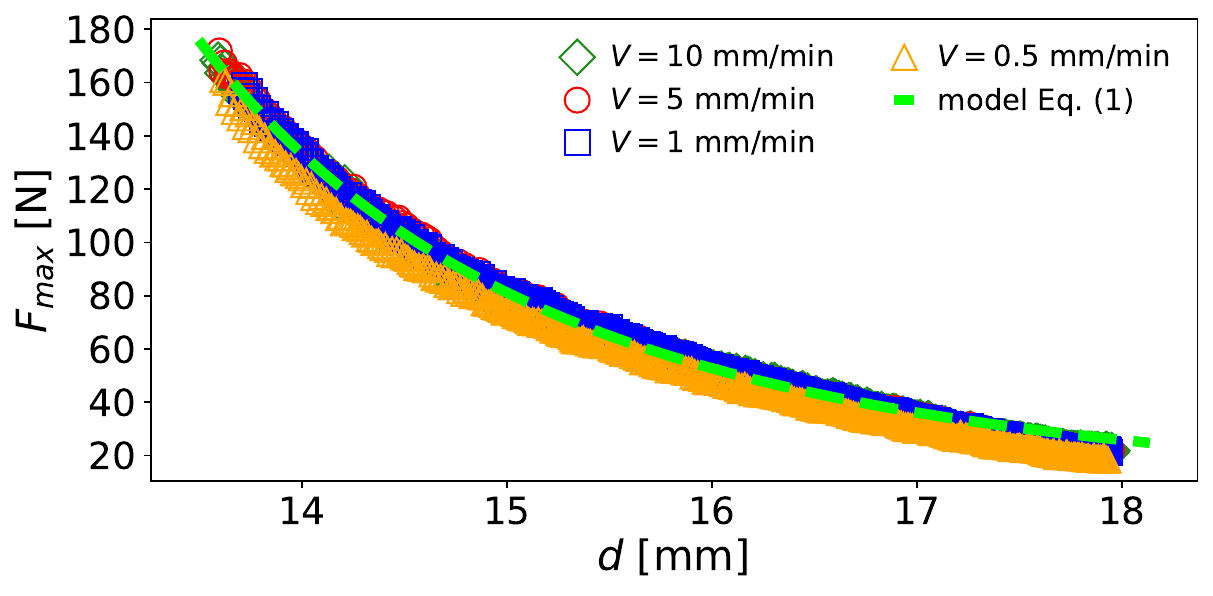}
    \caption{Maximum local force within single stick-slip events as a function of the separation distance for different pulling velocities. The dashed line shows the prediction of Eq.~\eqref{eq::model} obtained with $T^*=3$ mN and $\mu=0.89$. The effects of velocity are negligible for this observable.} 
    \label{fig:MaxVsD}
\end{figure}
The only control parameter that can be mapped directly from a simple spring-block configuration to our system is the traction velocity $V$, which acts as $v$ in the spring-block model. The effective stiffness of the interleaved assembly and the normal forces acting perpendicular to the sliding surfaces are both expected to depend non-trivially on the degree of entanglement of the two blocks (here quantified by $d$).  We therefore begin by comparing stick-slip features for different values of $V$ and $d$. First, we note that $F_{\text{max}}(d)$ does not substantially depend on $V$ for all the explored range of $d$ (Fig.~\ref{fig:MaxVsD}). A slight downward shift can be seen in the $V=0.5$ mm/min curve, but this can be attributed to aging, given that this curve corresponds to the last experiment performed in chronological order. Nevertheless, we are able to fit all the obtained $F_{\text{max}}(d)$ via Eq.~\eqref{eq::model}, using the same $T^*=3$ mN and $\mu=0.890 \pm 0.005$.
This is consistent with the scenario predicted by the spring-block model, where $f_\text{max}$ is independent of $v$ regardless of whether the low velocity limit is satisfied. We specify here that, throughout our study, every time we fit $F_\text{max}(d)$ via Eq.~\eqref{eq::model}, we must add an offset $\mathcal{T}_0$ to the latter, as our force signals are relative to $d=d_0$ (where the force sensor is set to zero). 
Across all our datasets, we obtained $\mathcal{T}_0\in[-10,10]$ N, which is a small interval compared to the full range of values found in a typical $F_\text{max}(d)$ curve.

We will now turn our attention to the force-displacement curve within individual stick-slip events. These events are identified by detecting local minima and maxima in $F(d)$. Each event begins at a local minimum $d=d_v$ and ends at the next local minimum, with the local maximum at $d=d_p$ occurring in between. It is also useful to define the local minimum force as $F_v = F(d_v)$ and the local maximum force as $F_p = F(d_p)$, which coincides with $F_{\text{max}}$ considered above (see also the two upper-right panels of Fig.~\ref{fig:StickSlip}).
\begin{figure*}
\includegraphics[width=0.99\textwidth]{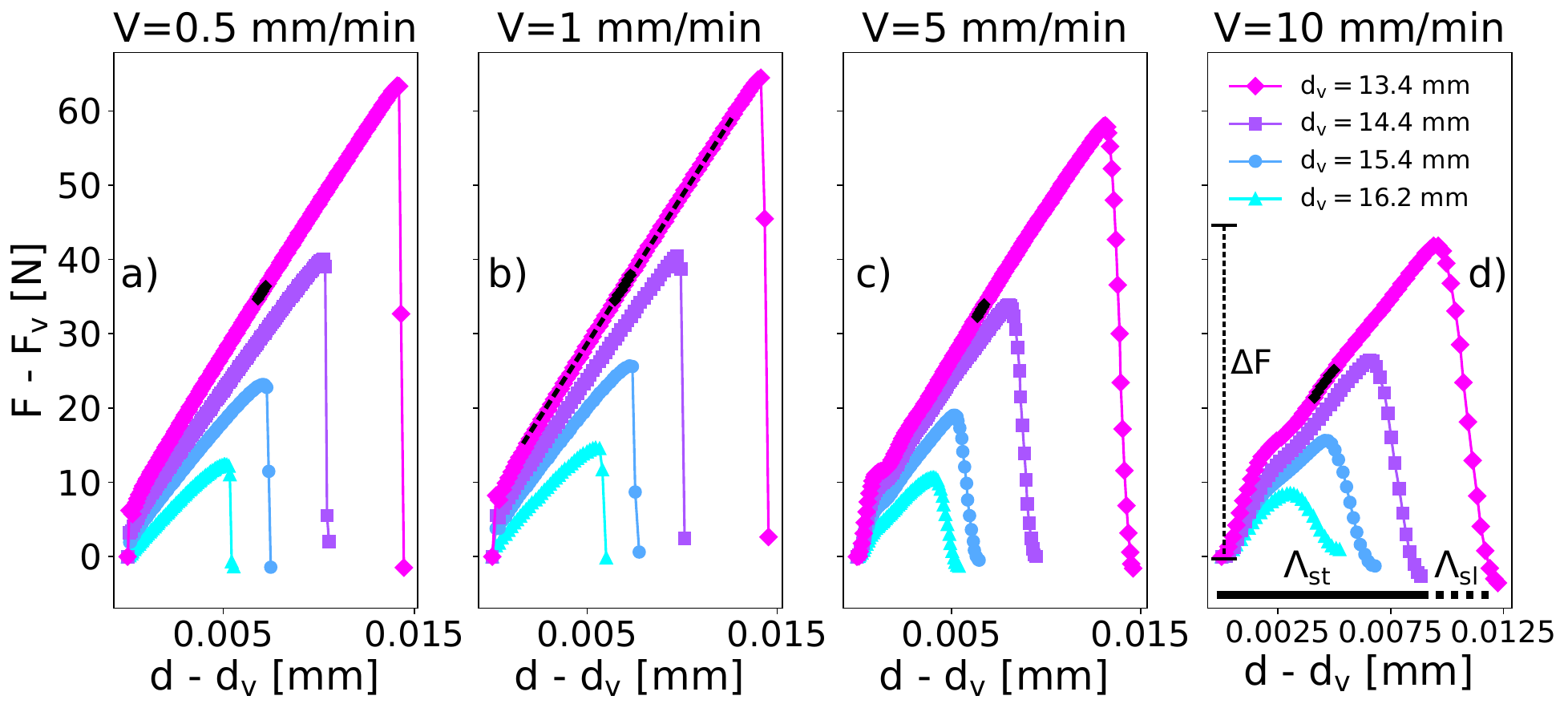} 
    \centering
    \caption{Force-displacement curves within single stick-slip events. Different panels show data obtained at different pulling velocities. Within each panel, we show stick-slip events starting at different separation distances $d_v$, which correspond to local force minima. In panel d, we define the stick-slip amplitude $\Delta F$ as well as the stick $\Lambda_{\text{st}}$ and slip $\Lambda_{\text{sl}}$ lengths. Each panel shows (on the pink curve) the subset of points used to estimate the effective stiffness of the assembly from the slope of the central part of the stick phase. Panel b) explicitly illustrates that this slope is representative of a significant portion of the stick.} 
    \label{fig:StickSlipEvents}
\end{figure*}
In Fig.~\ref{fig:StickSlipEvents}, we plot $F-F_v$ as a function of $d-d_v$ for different $d_v$ (in each panel) and $V$ (in different panels). The stick-slip amplitude $\Delta F=F_p-F_v$, the stick length $\Lambda_{\textbf{st}}=d_p-d_v$ and the slip length $\Lambda_{\textbf{sl}}$ are represented in Fig.~\ref{fig:StickSlipEvents}d. For the remainder of the paper, the stick-slip features defined here will be labelled as a function of $d_v$. 
From Fig.~\ref{fig:StickSlipEvents}, we observe that increasing $d_v$ at fixed $V$, or vice versa, has the same qualitative effect of reducing  $\Delta F$ and increasing $\Lambda_{\textbf{sl}}/\Lambda_{\textbf{st}}$ making the force-displacement curves appear more symmetrical. 
Nevertheless, the presence of a well-defined low velocity limit is indicated by the lack of significant change between the two lowest velocities, $V=0.5$ and $V=1$ mm/min.
For these two cases, we note that the slips are almost instantaneous (i.e. $\Lambda_{\textbf{sl}}/\Lambda_{\textbf{st}}\sim 0$) and that the stick phase is characterized by a linear increase in force, which is well described by a single slope. In contrast, at $V=5$ and $V=10$ mm/min, the stick exhibits a steep slope at small displacements, followed by a gentler slope.  
We do not have a clear explanation for the presence of this double regime at high velocities; the observed phenomenology suggests that it is a dynamical effect due to inertia. \modif{However, it is not simple to find a direct counterpart of this phenomenon in standard stick-slip models. A speculative interpretation for the observed  stiffness crossover within stick phases is briefly discussed at the end of Sec.~\ref{sec:effnom}.} \modif{A quantitative estimate of the velocity threshold below which the low-velocity limit holds would require considering additional velocities in the range $V\in [1,5]$ mm/min.} In what follows, we will mainly concentrate on the stick-slip features that survive in the low velocity regime \modif{$V \leq 1$ mm/min}.
\subsubsection{Stick-slip amplitude}
\begin{figure}
\includegraphics[width=0.99\columnwidth]{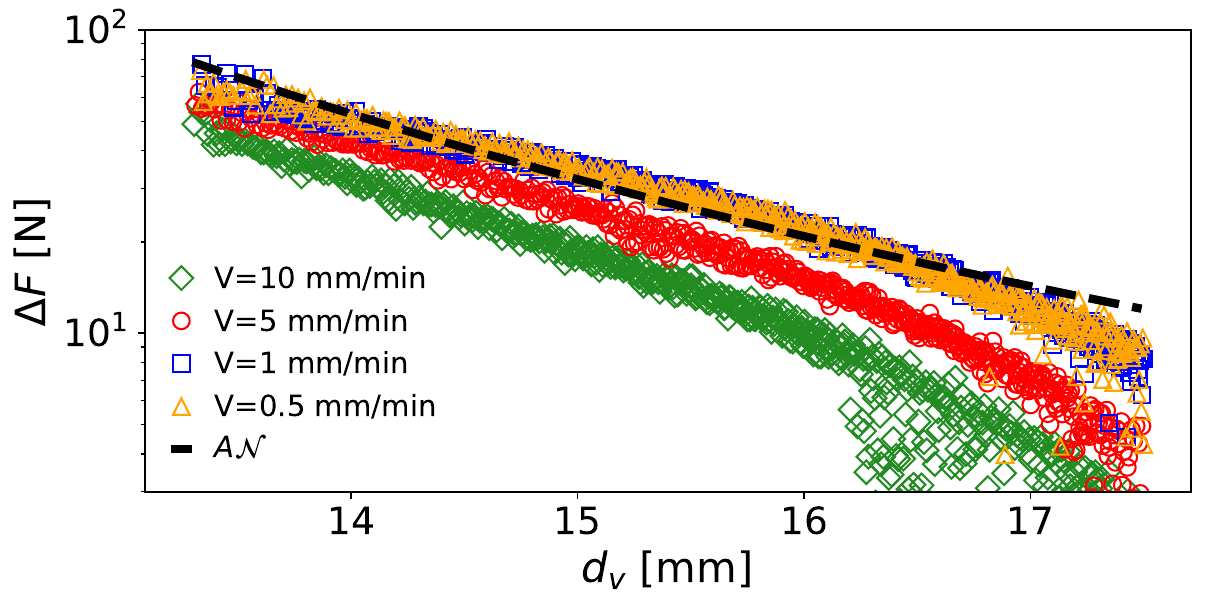}
    \caption{Stick-slip amplitude as a  function of the separation distance at the beginning of the stick for different pulling velocities. We note that the curves collapse at low velocities. The dashed line shows that the decay is relatively well approximated by the effective normal force acting on the assembly. This is proportional to the decay predicted by Eq.~\eqref{eq::model} with $T^*=3$ mN and $\mu=0.89$ through a prefactor $A/\mu$, with $A=0.34$.} 
    \label{fig:DropVsd}
\end{figure}
To make more quantitative considerations, we plot in Fig.~\ref{fig:DropVsd} the stick-slip amplitude of all the detected events as a function of $d_v$ for different velocities. Here we see that the values of $\Delta F$ stabilize at low velocities, independently of $d_v$. We point out that an increase in stick-slip amplitude until saturation has been observed in experiments  with sliding contact surfaces when the relative velocity decreases~\cite{kato1975stick}. In order to elucidate this phenomenon, it was necessary to employ a realistic friction model that incorporated a sticking-time-dependent friction coefficient~\cite{gao1994dynamic}. We also note that the curve at $V=10$ mm/min changes from exhibiting relatively small fluctuations to large ones around $d_v\sim 16.2$. We checked that this corresponds to the occurrence of less regular stick-slip events, which could be linked to a transition from stick-slip to steady sliding~\cite{Heslot1994} occurring in the outer region of the assembly, where normal force is low (see Eq.~\eqref{eq::normal}).  

Putting aside these dynamical effects arising at large $V$, we now focus on the two $\Delta F(d_v)$ curves in the low velocity regime with the aim of understanding their decrease during the detachment.  As discussed under Eq.~\eqref{eq::normal}, we expect the effective normal force $\mathcal{N}$  felt by the assembly to decrease proportionally to $\mathcal{T}$. As shown in Fig.~\ref{fig:DropVsd}, combining Eq.~\eqref{eq::model} and \eqref{eq::normal}, we verified that the range of values covered by our experimental data is compatible with $\Delta F(d_v)= A\mathcal{N}(d_v)$ with $A=0.34$. Following the analogy with the spring-block model, where the stick-slip amplitude decreases proportionally to the normal load with a prefactor $2(\mu_s-\mu_d)$, we can interpret $A$ as playing the role of this prefactor. 
As a further support of this interpretation, we point out that the estimated $A/2$ is very close to the value of $\mu_s-\mu_d$ obtained from our paper-on-paper friction characterization in the range of normal forces between 0.5 and 1 N, where $\mu_s\sim 0.61$ and $\mu_d\sim 0.44$ (see Appendix~\ref{app:singlesh}).
 We also note, however, that the proportionality between $\Delta F(d_v)$ and $\mathcal{N}(d_v)$ is not in perfect agreement with the data. 
 To better capture the functional form of this decay, it is probably necessary to improve the model proposed in Ref.~\cite{alarcon2016self} by taking into account stick-slip motion between contact surfaces. This is indeed a promising direction for future theoretical studies.
From our analysis of Fig.~\ref{fig:DropVsd}, we can nevertheless conclude that the decrease of the stick-slip amplitude as a function of $d_v$ is a consequence of the decrease in the effective normal force experienced by the contact surfaces in the assembly as it becomes less interleaved.
The reduction in stick-slip amplitude with normal force can then be easily understood based on an analogy with the spring-block model.

\subsubsection{Effective stiffness}\label{sec:ststic:stFeat:stiff}
\begin{figure}
\includegraphics[width=0.99\columnwidth]{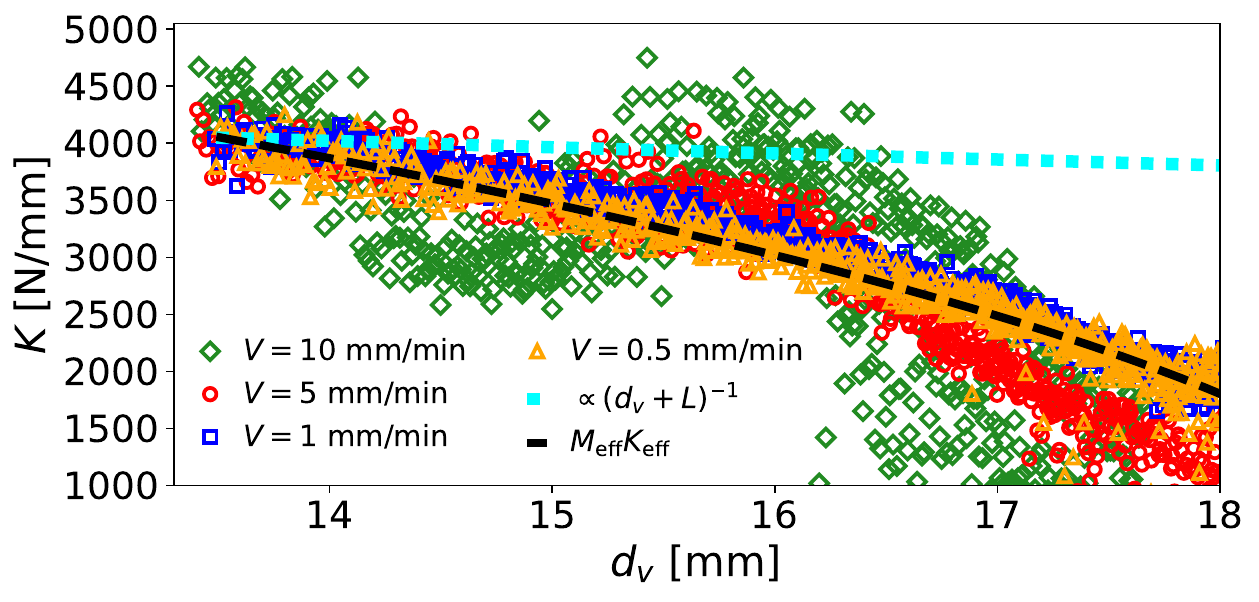}
    \caption{Effective stiffness of the assembly as a  function of the separation distance for different pulling
velocities. This is estimated from
the slope of the central part of the stick phase (see Fig.~\ref{fig:StickSlipEvents}). Also for this observable, the curves collapse at low velocities. The cyan dotted curve provides a comparison with the stiffness decay expected from the increase in assembly length during the detachment. The black dashed curve shows our prediction based on the effective number of pages contributing to the global stiffness (see Sec.~\ref{sec:effnom}).  Data for $V=5$ and $10$ mm/min reflect the presence of a double regime in the force-displacement curves during the stick.} 
    \label{fig:SlopeVSd}
\end{figure}
We now analyze the behavior of the slope of the force-displacement curve during the stick phase. In particular, we focus on the low velocity limit, where we expect this slope to represent the effective stiffness of the assembly.  By performing a linear fit of the central part of the force-displacement curve, we are sure to avoid possible non-linear effects due to the transition between stick and slip. This is, of course, only true for velocities where the stick phase is well described by a single linear regime.
In Fig.~\ref{fig:SlopeVSd}, we show the effective stiffness $K$ extracted from the central part of the force-displacement curve as a function of $d_v$ for different velocities. As expected, we observe that the data exhibit the same behavior for $V < 5$ mm/min, indicating a well-defined low velocity limit with no significant dynamic effects during the stick phase.  Focusing on this limit, we observe that $K$ monotonically decreases by a factor $\sim 4$ from the start to the end of the detachment. In other words, the assembly becomes softer as the degree of entanglement between the two blocks decreases. In order to better understand this behavior, we can first compare the measured values with the typical stiffness of the elementary elements of the assembly. Through traction tests performed on single paper sheets (see Appendix~\ref{app:singlesh}), we estimate a Young's modulus of the paper $Y_p\in [2,3]$ GPa. The stiffness of a single sheet under tensile stress would then be $K_p= \epsilon W Y_p/L\sim 200$ N/mm. Linking this elementary stiffness to that of the interleaved assembly is, however, not an easy task. By approximating the contribution of a couple of sheets with a series of two springs of stiffness $K_p$ and then considering the whole assembly as the equivalent of $2M$ of these springs in parallel, one can estimate the global stiffness to be $2M(K_p/2) \sim 5000$ N/mm, which matches the order of magnitude of the measured values. However, this approach lacks a consistent way to predict the observed decrease of $K$ as a function of $d_v$. By modelling the contribution of the individual sheets to the global stiffness more realistically, one could explain the decay of $K(d_v)$, considering that the assembly softens as its total length $L + d_v$ increases. In Fig.~\ref{fig:SlopeVSd}, we show that the expected stiffness decay $\propto(L+d_v)^{-1}$ appears almost flat in the explored range of $d_v$. In order to correctly understand the behavior of the global stiffness, it then becomes important to make more direct empirical observations of how the assembly deforms during detachment. This will be the main focus of the next section.

\section{The role of deformation}\label{sec:defo}
\subsection{Horizontal deformation by image processing}\label{sec:video}
\begin{figure}
\includegraphics[width=0.99\columnwidth]{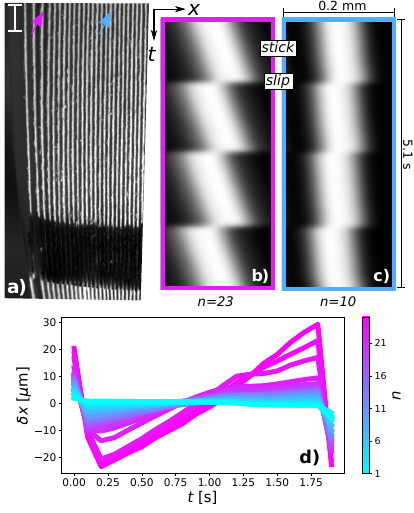}
    \caption{a) Image of a portion of the interleaved part of the assembly. The black-colored region highlights the bottom interleaved point. The white mark in the upper part of the image refers to the 0.84 mm vertical interval over which pixel intensity is averaged to perform image processing.  The colored arrows refer to the sheets used to produce the space-time diagram of panels b and c. Note that the external sheets are slightly bent outwards, whereas the internal ones are straight.
   b) Typical space-time diagram for an external sheet. It shows the time evolution of the pixel intensity in a 0.2 mm width horizontal interval centered around the $n=23$ sheet. \modif{Each row of the image corresponds to the light intensity averaged over the vertical interval shown in panel a. Different rows are obtained from consecutive frames and are stacked vertically, with the arrow of time pointing from top to bottom.} The considered time interval is \modif{5.1} seconds and corresponds to the occurrence of four stick-slip events. c) Same as panel b but for an internal sheet ($n=10$). It displaces significantly less than the outer sheet. d) Horizontal displacement between two slip events for all the sheets in the \modif{positive $x$} side of the assembly as a function of time. Outer sheets displace more than inner sheets. The data come from an event starting at $d=13.7$ mm in a traction test at $V = 1$ mm/min . } 
    \label{fig:ImageProc}
\end{figure}
To take a closer look at how the system deforms during the detachment, we performed some experiments during which we acquired high resolution images of the assembly from the frontal view (see Fig.~\ref{fig:ImageProc}a). By looking at the videos obtained from these images (see Supplementary Materials), we observed the following key facts: i) within the time resolution of our videos, slip events occur at the same time for all the sheets, ii) during both the stick and slip phases, the sheets move in the direction perpendicular to the traction force, iii) this horizontal motion is more significant for the outer sheets than the inner ones. These observations are summarized in  Figs.~\ref{fig:ImageProc}b-c where we show the space-time diagram during four stick-slip events for an external sheet ($n=23$, panel b) and a more internal one ($n=10$, panel c). These diagrams show the instantaneous pixel intensity averaged over a vertical interval of 0.84 mm in the upper part of the images (see the white mark in Fig.~\ref{fig:ImageProc}a). The horizontal axis represents the space interval around a sheet, the vertical axis represents the time evolution. It is clear that both sheets move inwards during the stick phase, until they experience an almost instantaneous outward jump in correspondence with the slip. This indicates that during the stick phase, when the assembly accumulates \modif{elastic potential energy}, it responds with horizontal compression; while it decompresses \modif{as such energy} is released in the slip. We also note that, during both the inward and outward motions, the external sheet moves across a wider horizontal interval than the internal one. 
As for the synchronicity of slip events, we believe it is true only at the timescale of our observation. An analogy with depinning models~\cite{jagla2014viscoelastic}, suggests a possible scenario where a single sheet reaches the threshold of static friction and then starts to slip. This alters the mechanical balance of the other sheets, triggering an avalanche of slip events. Whether the avalanche initiation occurs randomly within the assembly or in a specific weak region that can be predicted using a detailed model of the interleaved assembly is an interesting question for future studies that could be answered by imaging the system with a high-speed camera. 

We now focus on a detailed analysis of the horizontal system's deformation. From our images, we are able to detect the instantaneous horizontal position $x(n,t)$ of the $n$-th sheet in a given portion of the assembly (see Appendix~\ref{app:improc}). As we did for the space-time diagrams, we will consider the horizontal position of the sheets in the upper part of the images. Based on the symmetry of the system, we focused on the \modif{positive $x$} side of the assembly. 
Denoting as $t_v$ the starting time of the stick phases and with $t_{\text{st}}$ their duration (which are both independent of $n$), we analyzed the horizontal displacement $\delta x(n,t_v+t)=x(n,t_v+t)-\hat{x}(n)$, where $\hat{x}(n)=1/t_{\text{st}}\int_{0}^{t_{\text{st}}}x(n,t_v+t)dt$ is the average horizontal position of the sheet  within the stick.  In Fig.~\ref{fig:ImageProc}d, we plot $\delta x$ as a function  of $t\in[0,t_{\text{st}}]$ for $n\in[1,M]$ during a single stick-slip event. We note that this quantity  captures the observation that the outer sheets tend to displace more than the internal ones. 
\begin{figure}
\includegraphics[width=0.99\columnwidth]{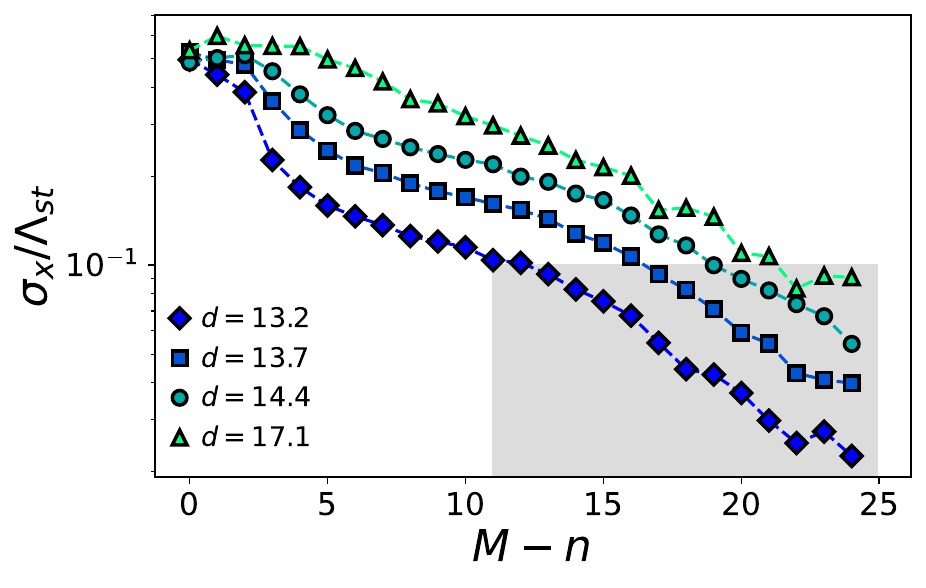}
    \caption{Mean absolute horizontal displacement during stick phases normalized by the stick length as a function of the page index. The data come from a traction test at $V=1$ mm/min. Each curve shows quantities averaged over \modif{12} stick phases starting at separation distances in the range $d\in[13.2,\modif{17.1}]$ mm. \modif{For all $d$}, outer sheets clearly displace more than the inner ones. \modif{The shaded gray rectangle contains the points below the threshold $\sigma_x/\Lambda_{\text{st}}<0.1$ which correspond to the internal shell of sheets effectively contributing to system's stiffness as estimated in Sec.~\ref{sec:effnom}. Their number varies between $14$ and $3$ throughout the detachment.}} 
    \label{fig:varPg}
\end{figure}
To measure the horizontal deformation of the system during the stick phases, we consider the mean absolute horizontal displacement 
$\sigma_x(n,t_v)=\sqrt{1/t_{\text{st}}\int_{0}^{t_{\text{st}}}\delta x^2(n,t_v+t)dt}$. As shown in Fig.~\ref{fig:varPg}, normalizing $\sigma_x(n,t_v)$ with the corresponding vertical stick length $\Lambda_{\text{st}}(t_v)=Vt_{\text{st}}(t_v)$, allows us to analyze the horizontal deformation of the assembly at different stages of detachment as a function of sheet index. \modif{The curves shown in Fig.~\ref{fig:varPg} provide a quantitative description of the scenario suggested by Fig.~\ref{fig:ImageProc}d. Indeed, $\sigma_x/\Lambda_{\text{st}}$ decreases as a function of $M-n$, meaning that the system's horizontal deformation reduces going from the external part to the internal part of the assembly.  Focusing on the numerical values of the vertical axis, we note that outer sheets always undergo horizontal displacements comparable to the vertical ones $\sigma_x/\Lambda_{\text{st}}\sim 0.5$. In contrast, the inner part of the assembly experiences an almost negligible horizontal deformation $\sigma_x/\Lambda_{\text{st}}<0.1$. By comparing curves at different $d$, we also observe that the detachment is accompanied by an increase in horizontal deformation across all the pages. 
This analysis suggests that the assembly could be considered as consisting of an outer shell with a high horizontal mobility and an inner shell of less mobile sheets. This scenario is preserved throughout the detachment but with an overall horizontal mobility that increases with the separation distance.} This picture will be particularly relevant for the next section.

One way to qualitatively explain  the observed behavior of the system's deformation perpendicular to the traction force  is to consider how the  force normal to the contact surfaces varies within the assembly. Even though the model developed in Ref.~\cite{alarcon2016self} does not take into account the deformability of the system, we can still use it to estimate the normal force $N(n)$ experienced by the sheets in the assembly using Eq.~\eqref{eq::normal}. 
As discussed in Sec.~\ref{sec:ststic:model}, we expect that the outer sheets experience a lower normal force, $N(n\sim M )\sim 0$, with respect to that felt by the inner ones $N(n\sim 1 )\gg 1$. Combined with the behavior of $\sigma_x/\Lambda_{\text{st}}$ discussed above, this reasonably suggests that the outer sheets, which experience a low $N$, are relatively more free to displace horizontally than the inner sheets, which experience a large $N$.

\subsection{Effective number of sheets}\label{sec:effnom}

Thanks to these insights into the system's horizontal deformation, we can now make an important step forward in understanding how the different sheets respond to the imposed \textit{vertical} traction. First, we recall that for each sheet $n$ in the assembly, we can define an angle $\theta(n,d)=\arctan{(n\epsilon/d)}$ between the vertical axis and the portion of the sheet connecting the clamping point to the start of the interleaved zone (see Fig.~\ref{fig:Setup}a). Each sheet under the static friction threshold can then respond to the imposed vertical displacement in two ways: by longitudinal elongation and/or by an inward horizontal displacement $\delta x(n,d)\le d\tan(\theta)= n\epsilon$. While the maximum possible horizontal displacement is purely determined by the geometry, the actual one also depends on the normal force $N(n,d)$ acting on the sheet.
The more the sheet is compressed \modif{along the $x-$axis}, the less it will be able to move. As a consequence, highly horizontally compressed sheets will mainly respond to imposed vertical motion through a longitudinal elongation of the order of the measured stick length $\Lambda_{\text{st}}$, thus with a stiffness $\propto \epsilon W Y_P/L$ where $Y_p$ is the Young's modulus of the paper (see also the discussion in Sec.~\ref{sec:ststic:stFeat:stiff}).

However, making an analogous estimate for less horizontally compressed sheets is not straightforward. From our images, we observed that they are slightly bent outwards (see again Fig.~\ref{fig:ImageProc}a). In this condition, despite the static friction constraint at the top interleaved point, these sheets can follow the imposed vertical motion without stretching by reducing their degree of bending (which involves local horizontal displacements).

As a crude approximation, we split the sheets in the assembly in two groups based on a horizontal compression threshold $\hat{N}$. Sheets identified by  $n\leq M_{\text{eff}}$ such that $N(n,d)\geq \hat{N}$ are considered not free to move \modif{along the $x$-axis}, thus responding to the vertical traction only through longitudinal stretching. The remaining less compressed sheets are instead considered as contributing with a negligible stiffness to the total effective one. Inverting Eq.~\eqref{eq::normal}, we can then obtain the following analytical expression for the number of sheets effectively contributing to the stiffness of the interleaved system:
\begin{equation}\label{eq::EffNumb}
    M_{\text{eff}}(d)=M\sqrt{1-\ln(1+2\mu \hat{N}/T^*)/\alpha},
\end{equation} which depends on $d$ through $\alpha$.
We can now come back to the problem of estimating the effective stiffness $K$ of the whole assembly introduced in Sec.~\ref{sec:ststic:stFeat:stiff}. Based on the above discussion, we consider $K$ as the result of $2M_{\text{eff}}$ couples of sheets acting as springs with stiffness ${K}_{\text{eff}}/2=\epsilon W {Y}_\text{eff}/(2L)$ in parallel, thus having $K(d)=M_{\text{eff}}(d){K}_{\text{eff}}$. Here, ${Y}_\text{eff}$ represents an effective Young modulus of the paper that we expect to be higher than the actual measured one $Y_p$ as it compensates for our crude approximation that considers the less compressed $M-M_{\text{eff}}$ sheets contributing with a zero stiffness to $K$. 
As shown in Fig.~\ref{fig:SlopeVSd}, using ${Y}_\text{eff}$ and $\hat{N}$ as fitting parameters, we obtain a good agreement between the measured $K(d_v)$ and the function $M_{\text{eff}}(d_v){K}_{\text{eff}}$ with ${Y}_\text{eff}=3.8$ GPa and $\hat{N}=0.57$ N.
The fitted value of  ${Y}_\text{eff}$ is then reasonable as it is close but slightly larger than the measured $Y_p$. Regarding the fitted value of $\hat{N}$, in Appendix~\ref{app:bend}, we discuss how we can link it with a typical confining normal force needed to bend a single sheet in the interleaved part of the assembly. 
\begin{figure}
\includegraphics[width=0.99\columnwidth]{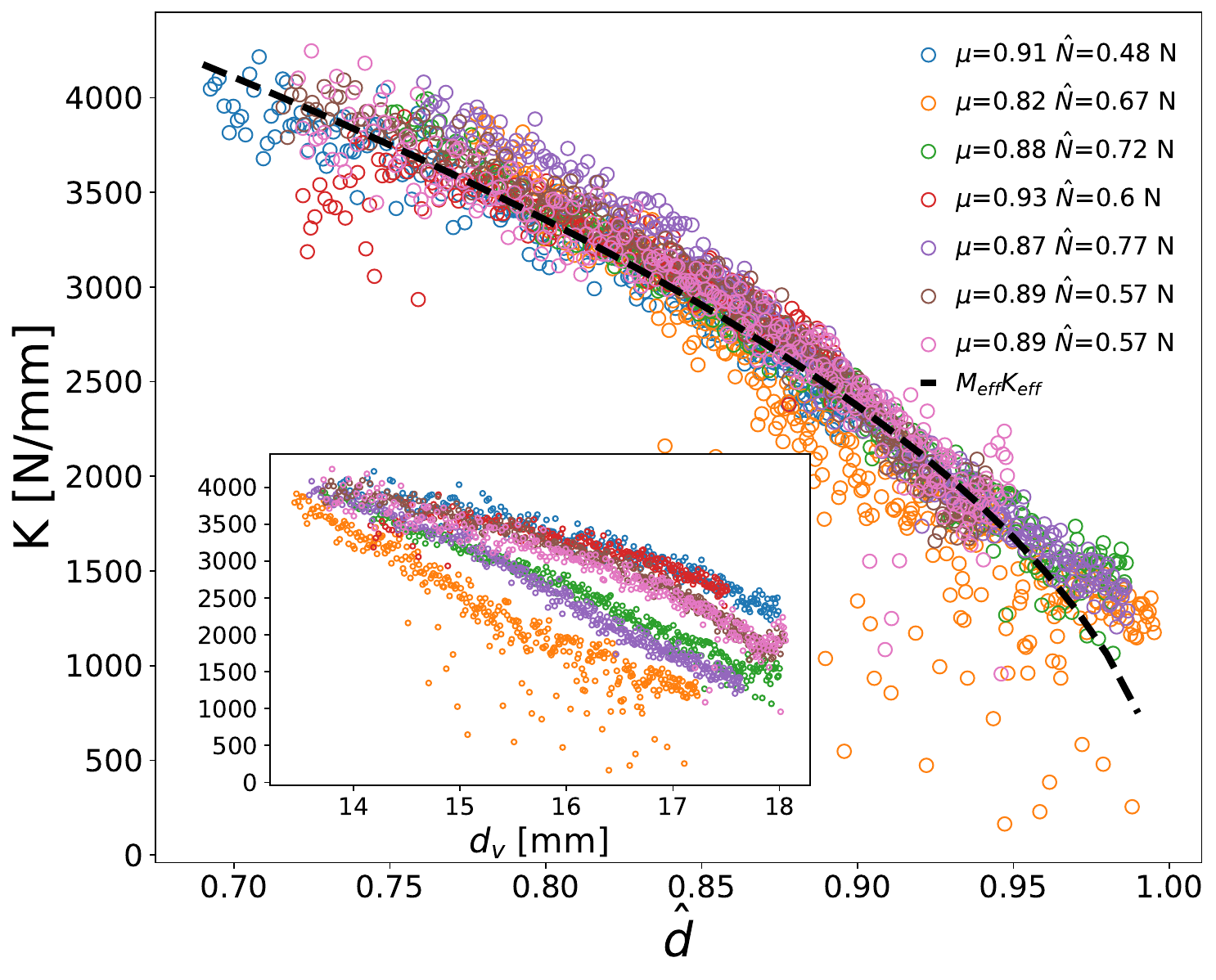}
    \caption{Effective stiffness of the assembly as a function of the adimensionalized separation distance $\hat{d}$ (defined in the text) for a series of traction tests at $V=1$ and $V=0.5$ mm/min. Data are obtained from samples
    differently aged, thus presenting a certain variability. This is shown in the inset, where the same data are plotted as a function of the separation distance at the beginning of the stick. Using $Y_\text{eff}=3.8$ GPa, $T^*=3$ mN, different $\mu$ fitted from the $F(d)$ curves and adjusting  $\hat{N}$, we are able to collapse all the data on the master curve predicted by our estimate based on the effective number of pages contributing to
    the global stiffness.} \label{fig:finalScaling}
\end{figure}

As a further test for the predictive power of our estimate of $K$, we now consider an additional series of experiments performed at $V=0.5$ or 1 mm/mm (i.e. in the low velocity limit). \modif{These experiments have been performed in a range of  ambient relative humidities between 28 $\%$ and 55 $\%$, with samples differently aged (i.e. between 9
and 28 preliminary traction tests in the range $d \in [13, 18]$ mm). The mechanical properties of our interleaved assemblies are expected to be impacted by humidity through its effect on the friction coefficient of the paper~\cite{borch2001handbook,kawashima2008paper}, as well as by various aging effects.} \modif{As discussed at the end of Sec.~\ref{sec:expset}}, we expect aging to occur at the level of the surface-contact asperities\modif{~\cite{de1997determination,garoff2004friction,fulleringer2014contribution}}. On the other hand, repeated traction protocols and sample manipulation can induce plastic deformation of the sheets, thereby altering their elastic response. This results in an observed variation of the force-displacement curves $F(d)$ and effective stiffness $K(d_v)$. Remarkably, for all the considered experiments, $F_{\text{max}}(d)$ can be fitted via Eq.~\eqref{eq::model} with the same $T^*=3$ mN as the one used before and slightly varying the friction coefficient in the range $\mu\in[0.82,0.93]$ (not shown). Keeping fixed these parameters along with ${Y}_\text{eff}=3.8$ GPa, we fitted the curves $K(d_v)$ with via $M_{\text{eff}}(d){K}_{\text{eff}}$ with only $\hat{N}$ as a free parameter. \modif{The effect of humidity and aging on the fitted $\mu$ and $\hat{N}$ is discussed in Appendix~\ref{app:fitPar}.} 
In Fig.~\ref{fig:finalScaling}, we show that this procedure allows us to collapse all the data in a plot as a function of the adimensional variable $\hat{d}=\ln(1+2\mu\hat{N}/T^*)/\alpha$.
As shown in the inset, the same data appear quite dispersed when plotted as a function of $d_v$.

We could now wonder about the possibility to establish a quantitative relation between the estimated $M_\mathrm{eff}$  
 and the system horizontal deformation measured through  image processing. 
 \modif{Starting from Fig.~\ref{fig:finalScaling}, we can check that $M_\mathrm{eff}=M\sqrt{1-\hat{d}}$ varies between 14 and 3 during the whole detachment where $\hat{d}\in[0.7,0.99]$. An analogous estimate of  $M_\mathrm{eff}$ based on the horizontal displacement of the sheets obtained through imaging is less straighthforward. However, looking at  Fig.~\ref{fig:varPg}, we note that the a similar variation in the number of sheets can be found considering the internal shell of pages where $\sigma_x/\Lambda_{\text{st}}< 0.1$ (see the points contained in the shaded area).  
 Of course, this quantitative link between the estimated  $M_\mathrm{eff}$ and sheets displacements is based on an empirical criterion. A direct independent estimate of  $M_\mathrm{eff}$ from imaging would require a proper reconstruction of the tensile strain field during the stick phases.  However, our combined analysis of Fig.~\ref{fig:varPg} and Fig.~\ref{fig:finalScaling} coherently validates the idea that the main contribution to the system's overall stiffness comes from highly compressed pages confined within an internal shell of the assembly.  }

\modif{The proposed mechanism based on an effective number of sheets participating in the system's stiffness also allows us to offer a tentative argument regarding the double stiffness regime observed for $V = 5$ and $10$ mm/min (i.e. beyond the low-velocity limit). We speculate that the initially higher stiffness may be due to a transient phase with a larger $M_{\mathrm{eff}}$. Considering that the relaxation toward the horizontal compression geometrically predicted by Eq.~\eqref{eq::normal} is not instantaneous, vertical detachment at high velocities may involve stick phases that begins while the $N(n)$ profile is still relaxing from the highly compressed state reached at the end of the previous stick phase. This would correspond to an initial higher compression and in turn to a higher transient $M_{\mathrm{eff}}$. However, this remains speculative; verifying it would require high-speed imaging to correctly resolve page displacement in the first instants of the stick phases at $V > 1$ mm/min.}

The agreement between the experimental data and our analytical estimate based on an effective number of sheets contributing to the global stiffness of the assembly suggests that, when examining the mechanical response of interleaved systems (manifesting here as stick-slip motion), the subtle interplay between frictional and elastic properties and internal stress distribution must be considered.  
In continuum mechanics, the response to body deformation is usually predicted by considering only the material's elastic constants. However, complex mechanical assemblies, such as the one considered here, also require an understanding of how the distribution of internal stresses depends on the frictional contact properties between the interleaved objects. Our analysis is built on the idea that the effective number of sheets contributing to the stiffness of the assembly (Eq.~\eqref{eq::EffNumb}) is determined by the distribution of normal forces within it (Eq.~\eqref{eq::normal}), which in turn depend on paper-on-paper friction. The system's response is entirely due to the coupling between traction and compression made possible by a finite friction coefficient between the sheets, a frictionless assembly would not exhibit any resistance to traction. This is consistent with the limit predicted by Eq.~\eqref{eq::EffNumb} for which $M_{\text{eff}}\to0$ when $\mu\to 0$.
Taking all these ingredients into account, it is reasonable to find that, in the presence of such an emergent friction-driven cohesion,  body deformation is governed by an effective stiffness that also depends on frictional properties.

\section{Future outlook}\label{sec:outlook}
\modif{While conducting this study, we identified several interesting aspects that could be explored further in future research. This section provides an overview of those perspectives.}
\\ \\
\textit{Stick-slip instability analysis. }When dealing with friction dynamics, a classic analysis concerns the transition between stick-slip and steady sliding. Empirical observations generally show that steady sliding becomes unstable with respect to stick-slip at high velocities, high normal forces, and low stiffness. Linear stability analyses of spring-block models with a rate and state-variable description, predict the instability of steady sliding when the ratio of spring stiffness to normal load is below a threshold that depends on the imposed velocity~\cite{Heslot1994}. Extending this type of analysis to interleaved systems is a valuable direction for future studies. However, suitable experimental protocols must first be designed, as the only control parameter that can be directly mapped into standard spring-block models  is $V$. As demonstrated by the investigation in this paper, in an interleaved assembly experiencing the self-amplification of frictional forces, the effective normal load and stiffness relevant to stick-slip dynamics are complex emergent properties of the system.
\\ \\
\textit{Theoretical modeling.} From a theoretical standpoint, we emphasize that, while the mechanical model proposed in Ref.~\cite{alarcon2016self} was useful for interpreting certain experimental results presented in this study, a more refined version of it that explicitly describes stick-slip motion would be a valuable tool for describing interleaved books and, more generally, for understanding the dynamics of systems with friction amplification. In this perspective, we identify three key aspects that should be considered in refining the model. First,  a proper description of sheet elastic deformation combined with dynamic equations for paper-on-paper contact surfaces that are able to describe stick-slip motion. 
This could be achieved by modelling each couple of sheets as a spring-block system, possibly including a rate-and state-variable framework to perform local linear stability analyses. 
Second, one should consider the fact that the normal load increases during the stick phase. Indeed, as the traction force grows, so does its normal projection acting on the interleaved part. Third, it would be desirable also to properly include in the model  \modif{a $\mu(N)$ dependence} of the static friction to the normal force due to adhesion (see also Appendix~\ref{app:singlesh}). 
\\ \\
\modif{\textit{Application oriented studies.} With regard to practical applications of interleaved book-like systems~\cite{an2024energy,luo2024switchable,Huang2021-qa}, it would be interesting to investigate the effect of the assembly's orientation on its mechanical response. For instance, rotating the system by ninety degrees in the $xz$-plane and pulling the interleaved blocks \modif{along the $x$-axis} would introduce an additional variation in normal stress due to the accumulated sheet weight. This force gradient would break the symmetry with respect to the pulling axis potentially impacting the stick-slip behavior. Of particular interest for applications would also be an investigation into the mechanical response of the assembly under repeated pull-push cycles, as such loading conditions are likely encountered in tactile displays~\cite{an2024energy} and switchable clutches~\cite{luo2024switchable}. This would reveal hysteretic effects reflecting irreversible mechanisms in both the bulk paper material (already evidenced for single sheets in Appendix A) and the contact dynamics, promising rich phenomenology. Fatigue effects may also emerge, manifesting as a progressive weakening of the resistance to detachment after repeated pull-push cycles below the slip condition.}
\\ \\
\modif{\textit{High-speed imaging.}} We also highlight the intriguing perspective of using high-speed imaging \modif{to properly investigate the mechanical response during stick phases at high detachment velocities and} to analyze the spatiotemporal distribution of slip avalanches within the assembly.
\\ \\
\modif{\textit{Beyond book-like systems.} We believe that our findings are relevant to a broader class of complex mechanical systems in which externally applied forces lead to a strengthening of internal frictional contacts. For example, in dense granular systems under shear, one would expect any contact surface between two grains with a normal vector that is not perpendicular to the applied shear to experience a frictional force effectively amplified by the shear itself.
A similar situation may arise in fibrous systems under tensile loading, where fibers can experience relative compression in the direction perpendicular to the applied tensile stress. In such systems, it is possible to identify variables analogous to the degree of interleaving, quantified in our case by the dimensionless parameter $\alpha$. In granular materials, this could correspond to the inverse of the volume fraction (i.e. more compact $\leftrightarrow$ more interleaved), or to more complex observables related to the contact network. In twisted yarns, a dimensionless parameter analogous to $\alpha$ could be defined based on the twist angle\cite{seguin2022twist}.
A natural extension of this study would then be to examine how mechanisms analogous to those identified in our study, namely the coupling between the evolution of the contact force distribution and the resulting stick-slip dynamics, can provide insight into the mechanical response of granular and fibrous materials near yielding, where such systems are known to exhibit stick slip behavior\cite{poincloux2018crackling,Hu2023-mq}.}

\section{CONCLUSION}\label{sec:conc}

In this paper, we studied stick-slip motion in a system formed from two interleaved blocks of paper sheets. This system exhibits unusual friction amplification against block separation due to the conversion of traction force into compression, which increases friction between the contact surfaces within the assembly. Our force measurements enable us to resolve individual stick-slip events and study different dynamic features as a function of the imposed detachment velocity and the assembly's degree of entanglement. The stick-slip amplitude decreases as the system detaches and as the pulling velocity increases. We demonstrated that this behavior is linked to the overall normal compression experienced by the contact surfaces in the assembly and can be understood through an analogy with a simple frictional spring-block model. By combining force measurements with an imaging-based analysis of the system's normal deformation, we analyzed the behavior of the assembly's global stiffness, estimated in the low velocity limit from the slope of the force-displacement curve in the stick phase. We found that this stiffness also decreases as the system detaches. This behavior can be explained by observing that the outer shell of the assembly deforms more than the inner shell in the direction \modif{perpendicular to the traction force} during the stick phase. Consequently, the outer shell contributes less to the overall effective stiffness measured under vertical traction. We then analytically estimated the effective number of sheets contributing to the global stiffness, taking into account the variation of the local confinement due to normal forces in the assembly. The good agreement between the stiffness obtained through this estimation and the experimental data highlights the non-trivial interplay between the internal stress distribution and the mechanical properties of the system. Indeed, the number of sheets that effectively contribute to the stiffness of the assembly depends on how normal forces are distributed within it, which in turn depends on the friction between paper surfaces.

To the best of our knowledge, this is the first systematic study of stick-slip motion in a complex mechanical assembly involving multiple surfaces in contact and friction amplification properties. Interleaved books represent a model system for this type of assemblies, meaning that the analysis reported here is also relevant to granular, textile and fibrous materials, as well as metamaterials that exploit friction amplification mechanisms.

\section*{DATA AVAILABILITY}
The raw data for the force measurements can be found at https://zenodo.org/record/18596578.

\begin{acknowledgments}

The authors acknowledge Sandrine Mariot for technical support in the use of the SoMaC\&CoMic platform. This work has been funded by the Agence Nationale
de la Recherche (ANR), France, grant ANR-21-CE06-0039.

\end{acknowledgments}

\bibliographystyle{ieeetr} 
\bibliography{biblio}

@article{Restagno2004,
  title = {Where does a cohesive granular heap break?},
  volume = {14},
  ISSN = {1292-895X},
  url = {http://dx.doi.org/10.1140/epje/i2004-10013-5},
  DOI = {10.1140/epje/i2004-10013-5},
  number = {2},
  journal = {The European Physical Journal E},
  publisher = {Springer Science and Business Media LLC},
  author = {Restagno,  F. and Bocquet,  L. and Charlaix,  E.},
  year = {2004},
  month = jun,
  pages = {177–183}
}

@article{shewan2020tribology,
  title={Tribology and its growing use toward the study of food oral processing and sensory perception},
  author={Shewan, Heather M and Pradal, Clementine and Stokes, Jason R},
  journal={Journal of texture studies},
  volume={51},
  number={1},
  pages={7--22},
  year={2020},
  publisher={Wiley Online Library}
}

@article{Duran98,
  title = {Static friction and arch formation in granular materials},
  author = {Duran, J. and Kolb, E. and Vanel, L.},
  journal = {Phys. Rev. E},
  volume = {58},
  issue = {1},
  pages = {805--812},
  numpages = {0},
  year = {1998},
  month = {Jul},
  publisher = {American Physical Society},
  doi = {10.1103/PhysRevE.58.805},
  url = {https://link.aps.org/doi/10.1103/PhysRevE.58.805}
}

@ARTICLE{Allan2025-fp,
  title     = "{Micro-CT} analysis of geometrical distribution of filaments in
               double-twisted three-strand ropes: Comparison with analytical
               model",
  author    = "Allan, Oday and Mishra, Tanmaya and Meijer, Robert Jan and
               Rooij, Matthijn de",
  journal   = "J. Ind. Text.",
  publisher = "SAGE Publications",
  volume    =  55,
  number    =  15280837251333116,
  month     =  apr,
  year      =  2025,
  language  = "en"
}

@article{armstrong2002stick,
  title={Stick slip and control in low-speed motion},
  author={Armstrong-Helouvry, Brian},
  journal={IEEE Transactions on Automatic Control},
  volume={38},
  number={10},
  pages={1483--1496},
  year={2002},
  publisher={IEEE}
}

@ARTICLE{Allan2025-dq,
  title     = "Experimental analysis of contact forces between strands in
               three-strand ropes with varying twist parameters and filament
               counts",
  author    = "Allan, Oday and Mishra, Tanmaya and Rooij, Matthijn de",
  journal   = "Text. Res. J.",
  publisher = "SAGE Publications",
  number    =  00405175251333381,
  month     =  may,
  year      =  2025,
  language  = "en"
}

@ARTICLE{Hu2023-mq,
  title     = "High time-resolved studies of stick-slip show similar dilatancy
               to fast and slow earthquakes",
  author    = "Hu, Wei and Ge, Yi and Xu, Qiang and Huang, Runqiu and Zhao, Qi
               and Gou, Huaixiao and McSaveney, Mauri and Chang, Chingshung and
               Li, Yan and Jia, Xiaoping and Wang, Yujie",
  journal   = "Proc. Natl. Acad. Sci. U. S. A.",
  publisher = "Proceedings of the National Academy of Sciences",
  volume    =  120,
  number    =  47,
  pages     = "e2305134120",
  month     =  nov,
  year      =  2023,
  copyright = "https://creativecommons.org/licenses/by-nc-nd/4.0/",
  language  = "en"
}

@BOOK{Bhushan2001-wh,
  title     = "Fundamentals of tribology and bridging the gap between the
               macro- and micro/nanoscales",
  editor    = "Bhushan, Bharat",
  publisher = "Springer",
  series    = "NATO Science Series II",
  month     =  "mar",
  year      =  "2001",
  address   = "Dordrecht, Netherlands",
  language  = "en"
}

@article{Rastei2013puck,
  title = {Puckering Stick-Slip Friction Induced by a Sliding Nanoscale Contact},
  author = {Rastei, M. V. and Heinrich, B. and Gallani, J. L.},
  journal = {Phys. Rev. Lett.},
  volume = {111},
  issue = {8},
  pages = {084301},
  numpages = {5},
  year = {2013},
  month = {Aug},
  publisher = {American Physical Society},
  doi = {10.1103/PhysRevLett.111.084301},
  url = {https://link.aps.org/doi/10.1103/PhysRevLett.111.084301}
}

@article{BaldassarriBrownian,
  title = {Brownian Forces in Sheared Granular Matter},
  author = {Baldassarri, A. and Dalton, F. and Petri, A. and Zapperi, S. and Pontuale, G. and Pietronero, L.},
  journal = {Phys. Rev. Lett.},
  volume = {96},
  issue = {11},
  pages = {118002},
  numpages = {4},
  year = {2006},
  month = {Mar},
  publisher = {American Physical Society},
  doi = {10.1103/PhysRevLett.96.118002},
  url = {https://link.aps.org/doi/10.1103/PhysRevLett.96.118002}
}

@INPROCEEDINGS{Huang2021-qa,
  title           = "Intersecting book inspired high-power-density
                     electret/triboelectric multilayered power generator with
                     flexible interdigital electrodes",
  booktitle       = "2021 21st International Conference on {Solid-State}
                     Sensors, Actuators and Microsystems (Transducers)",
  author          = "Huang, Hao and Zhao, Zhe and Tao, Kai and Wu, Jin and Ji,
                     Bowen and Yuan, Weizheng and Chang, Honglong",
  publisher       = "IEEE",
  month           =  jun,
  year            =  2021,
  copyright       = "https://ieeexplore.ieee.org/Xplorehelp/downloads/license-information/IEEE.html",
  conference      = "2021 21st International Conference on Solid-State Sensors,
                     Actuators and Microsystems (Transducers)",
  location        = "Orlando, FL, USA"
}

@article{poincloux2024stick,
  title={Stick-slip in a stack: How slip dissonance reveals aging},
  author={Poincloux, Samuel and Reis, Pedro M and de Geus, Tom WJ},
  journal={Physical Review Research},
  volume={6},
  number={1},
  pages={013080},
  year={2024},
  publisher={APS}
}

@article{poincloux2021bending,
  title={Bending response of a book with internal friction},
  author={Poincloux, Samuel and Chen, Tian and Audoly, Basile and Reis, Pedro M},
  journal={Physical Review Letters},
  volume={126},
  number={21},
  pages={218004},
  year={2021},
  publisher={APS}
}

@article{luo2024switchable,
  title={A switchable flexible mechanical clutch based on self-amplified friction of interleaved layers},
  author={Luo, Aoyi and Hart, A John},
  journal={Extreme Mechanics Letters},
  volume={71},
  pages={102197},
  year={2024},
  publisher={Elsevier}
}

@article{an2024energy,
  title={Energy-efficient dynamic 3D metasurfaces via spatiotemporal jamming interleaved assemblies for tactile interfaces},
  author={An, Siqi and Li, Xiaowen and Guo, Zengrong and Huang, Yi and Zhang, Yanlin and Jiang, Hanqing},
  journal={Nature Communications},
  volume={15},
  number={1},
  pages={7340},
  year={2024},
  publisher={Nature Publishing Group UK London}
}

@UNPUBLISHED{Jiang2025-lk,
  title  = "Emergent neural network-like mechanical response in interlocking
            materials",
  author = "Jiang, Peng and Qi, Jixiang and Shi, Zengqin and Yang, Heng and Li,
            Ying",
  year   =  2025
}

@article{wierzchalek2025vane,
  title={Vane rheology of a fiber-reinforced granular material},
  author={Wierzchalek, Ladislas and Gauthier, Georges and Darbois Texier, Baptiste},
  journal={Journal of Rheology},
  volume={69},
  number={3},
  pages={353--363},
  year={2025},
  publisher={AIP Publishing}
}

@article{vani2025asymmetric,
  title={Asymmetric bending boundary layer: The $\lambda$-test},
  author={Vani, Nathan and Ibarra, Alejandro and Bico, Jos{\'e} and Reyssat, {\'E}tienne and Roman, Beno{\^\i}t},
  journal={Proceedings of the National Academy of Sciences},
  volume={122},
  number={11},
  pages={e2426748122},
  year={2025},
  publisher={National Academy of Sciences}
}

@article{weiner2020mechanics,
  title={Mechanics of randomly packed filaments—the “bird nest” as meta-material},
  author={Weiner, Nicholas and Bhosale, Yashraj and Gazzola, Mattia and King, Hunter},
  journal={Journal of Applied Physics},
  volume={127},
  number={5},
  year={2020},
  publisher={AIP Publishing}
}

@article{franklin2012geometric,
  title={Geometric cohesion in granular materials},
  author={Franklin, Scott V},
  journal={Physics today},
  volume={65},
  number={9},
  pages={70--71},
  year={2012},
  publisher={AIP Publishing}
}

@article{lu2025experimental,
  title={Experimental investigation of stick-slip behaviors in dry sliding friction},
  author={Lu, Yixiao and Han, Dong and Fu, Qidi and Lu, Xi and Zhang, Yan and Wei, Zhiyong and Chen, Yunfei},
  journal={Tribology International},
  volume={201},
  pages={110221},
  year={2025},
  publisher={Elsevier}
}

@article{dong2017stick,
  title={Stick-slip behaviours of water lubrication polymer materials under low speed conditions},
  author={Dong, Conglin and Shi, Lichun and Li, Lvzhou and Bai, Xiuqin and Yuan, Chengqing and Tian, Yu},
  journal={Tribology International},
  volume={106},
  pages={55--61},
  year={2017},
  publisher={Elsevier}
}

@article{zhang2023influence,
  title={The influence of interfacial wear characteristics on stick-slip vibration},
  author={Zhang, QX and Mo, JL and Xiang, ZY and Liu, QA and Tang, Bin and Jin, WW and Zhu, Song},
  journal={Tribology International},
  volume={185},
  pages={108535},
  year={2023},
  publisher={Elsevier}
}

@article{achanta2010scale,
  title={On the scale dependence of coefficient of friction in unlubricated sliding contacts},
  author={Achanta, Satish and Celis, J-P},
  journal={Wear},
  volume={269},
  number={5-6},
  pages={435--442},
  year={2010},
  publisher={Elsevier}
}

@article{dalbe2015multiscale,
  title={Multiscale stick-slip dynamics of adhesive tape peeling},
  author={Dalbe, Marie-Julie and Cortet, Pierre-Philippe and Ciccotti, Matteo and Vanel, Lo{\"\i}c and Santucci, St{\'e}phane},
  journal={Physical review letters},
  volume={115},
  number={12},
  pages={128301},
  year={2015},
  publisher={APS}
}

@article{dumont2018emergent,
  title={Emergent strain stiffening in interlocked granular chains},
  author={Dumont, Denis and Houze, Maurine and Rambach, Paul and Salez, Thomas and Patinet, Sylvain and Damman, Pascal},
  journal={Physical review letters},
  volume={120},
  number={8},
  pages={088001},
  year={2018},
  publisher={APS}
}

@book{schwartz2019structure,
  title={Structure and mechanics of textile fibre assemblies},
  author={Schwartz, Peter},
  year={2019},
  publisher={Woodhead publishing}
}

@article{poincloux2018crackling,
  title={Crackling dynamics in the mechanical response of knitted fabrics},
  author={Poincloux, Samuel and Adda-Bedia, Mokhtar and Lechenault, Fr{\'e}d{\'e}ric},
  journal={Physical review letters},
  volume={121},
  number={5},
  pages={058002},
  year={2018},
  publisher={APS}
}

@article{poincloux2018geometry,
  title={Geometry and elasticity of a knitted fabric},
  author={Poincloux, Samuel and Adda-Bedia, Mokhtar and Lechenault, Fr{\'e}d{\'e}ric},
  journal={Physical Review X},
  volume={8},
  number={2},
  pages={021075},
  year={2018},
  publisher={APS}
}

@article{brown2015physiological,
  title={The physiological molecular shape of spectrin: a compact supercoil resembling a Chinese finger trap},
  author={Brown, Jeffrey W and Bullitt, Esther and Sriswasdi, Sira and Harper, Sandra and Speicher, David W and McKnight, C James},
  journal={PLoS computational biology},
  volume={11},
  number={6},
  pages={e1004302},
  year={2015},
  publisher={Public Library of Science San Francisco, CA USA}
}

@article{su2012modified,
  title={The modified finger-trap suture technique: a biomechanical comparison of a novel suture technique for graft fixation},
  author={Su, Wei-Ren and Chu, Chun-Hui and Lin, Cheng-Li and Lin, Chii-Jen and Jou, I-Ming and Chang, Chih-Wei},
  journal={Arthroscopy: The Journal of Arthroscopic \& Related Surgery},
  volume={28},
  number={5},
  pages={702--710},
  year={2012},
  publisher={Elsevier}
}

@article{ghosal2012capstan,
  title={Capstan friction model for DNA ejection from bacteriophages},
  author={Ghosal, Sandip},
  journal={Physical review letters},
  volume={109},
  number={24},
  pages={248105},
  year={2012},
  publisher={APS}
}

@article{jung2008generalized,
  title={Generalized capstan problem: Bending rigidity, nonlinear friction, and extensibility effect},
  author={Jung, Jae Ho and Pan, Ning and Kang, Tae Jin},
  journal={Tribology International},
  volume={41},
  number={6},
  pages={524--534},
  year={2008},
  publisher={Elsevier}
}

@article{leech2002modelling,
  title={The modelling of friction in polymer fibre ropes},
  author={Leech, CM},
  journal={International Journal of Mechanical Sciences},
  volume={44},
  number={3},
  pages={621--643},
  year={2002},
  publisher={Elsevier}
}

@article{crassous1999humidity,
  title={Humidity effect on static aging of dry friction},
  author={Crassous, J and Bocquet, L and Ciliberto, S and Laroche, C},
  journal={Europhysics Letters},
  volume={47},
  number={5},
  pages={562},
  year={1999},
  publisher={IOP Publishing}
}

@article{dreier2025beaded,
  title={Beaded metamaterials},
  author={Dreier, Lauren and Jones, Trevor J and Plummer, Abigail and Ko{\v{s}}mrlj, Andrej and Brun, P-T},
  journal={Nature communications},
  volume={16},
  number={1},
  pages={7899},
  year={2025},
  publisher={Nature Publishing Group UK London}
}

@phdthesis{taub2020assemblages,
  title={Assemblages d'objets {\'e}lanc{\'e}s: m{\'e}canique et effets de contact},
  author={Taub, Raphaelle},
  year={2020},
  school={Universit{\'e} Paris-Saclay}}

@article{taub2021nonlinear,
  title={Nonlinear amplification of adhesion forces in interleaved books},
  author={Taub, Raphaelle and Salez, Thomas and Alarcon, Hector and Rapha{\"e}l, {\'E}lie and Poulard, Christophe and Restagno, Fr{\'e}d{\'e}ric},
  journal={The European Physical Journal E},
  volume={44},
  number={5},
  pages={71},
  year={2021},
  publisher={Springer}
}

@article{alarcon2016self,
  title={Self-amplification of solid friction in interleaved assemblies},
  author={Alarc{\'o}n, H{\'e}ctor and Salez, Thomas and Poulard, Christophe and Bloch, Jean-Francis and Rapha{\"e}l, {\'E}lie and Dalnoki-Veress, Kari and Restagno, Fr{\'e}d{\'e}ric},
  journal={Physical review letters},
  volume={116},
  number={1},
  pages={015502},
  year={2016},
  publisher={APS}
}

@article{seguin2022twist,
  title={Twist-controlled force amplification and spinning tension transition in yarn},
  author={Seguin, Antoine and Crassous, J{\'e}r{\^o}me},
  journal={Physical Review Letters},
  volume={128},
  number={7},
  pages={078002},
  year={2022},
  publisher={APS}
}

@article{elmer1997nonlinear,
  title={Nonlinear dynamics of dry friction},
  author={Elmer, Franz-Josef},
  journal={Journal of Physics A: Mathematical and General},
  volume={30},
  number={17},
  pages={6057},
  year={1997},
  publisher={IOP Publishing}
}

@article{gao1993fundamentals,
  title={Fundamentals of stick-slip},
  author={Gao, Chao and Kuhlmann-Wilsdorf, Doris and Makel, David D},
  journal={Wear},
  volume={162},
  pages={1139--1149},
  year={1993},
  publisher={Elsevier}
}

@article{kato1975stick,
  title={STICK-SLIP MOTION AND CHARACTERISTICS OF FRICTION IN MACHINE TOOL, SLIDEWAY},
  author={Kato, SHINOBU and Yamaguchi, KATsUy and Matsubayashi, Tsuneo and Sato, Norio},
  journal={Memoirs of the Faculty of Engineering, Nagoya University},
  volume={27},
  number={1},
  pages={1--71},
  year={1975},
  publisher={Faculty of Engineering, Nagoya University}
}

@article{gao1994dynamic,
  title={The dynamic analysis of stick-slip motion},
  author={Gao, Chao and Kuhlmann-Wilsdorf, Doris and Makel, David D},
  journal={Wear},
  volume={173},
  number={1-2},
  pages={1--12},
  year={1994},
  publisher={Elsevier}
}

@article{plati2025control,
  title={Control of friction: Shortcuts and optimization for the rate-and state-variable equation},
  author={Plati, Andrea and Petri, Alberto and Baldovin, Marco},
  journal={European Journal of Mechanics-A/Solids},
  volume={111},
  pages={105550},
  year={2025},
  publisher={Elsevier}
}

@article{jagla2014viscoelastic,
  title={Viscoelastic effects in avalanche dynamics: A key to earthquake statistics},
  author={Jagla, Eduardo Alberto and Landes, Fran{\c{c}}ois P and Rosso, Alberto},
  journal={Physical review letters},
  volume={112},
  number={17},
  pages={174301},
  year={2014},
  publisher={APS}
}

@article{Baumberger1996,
  title = {Creeplike Relaxation at the Interface between Rough Solids under Shear},
  volume = {6},
  ISSN = {1286-4862},
  url = {http://dx.doi.org/10.1051/jp1:1996113},
  DOI = {10.1051/jp1:1996113},
  number = {8},
  journal = {Journal de Physique I},
  publisher = {EDP Sciences},
  author = {Baumberger,  T. and Gauthier,  L.},
  year = {1996},
  pages = {1021–1030}
}

@book{Bowden1950,
author = {Bowden, F. P.  and Tabor, D.},
title = {The Friction and Lubrication of Solids},
year= {1950},
publisher = {Clarendon Press},
location = {London}
}

@article{Vanossi2013,
	title = {Colloquium: Modeling friction: From nanoscale to mesoscale},
	author = {Vanossi, Andrea and Manini, Nicola and Urbakh, Michael and Zapperi, Stefano and Tosatti, Erio},
	journal = {Rev. Mod. Phys.},
	volume = {85},
	issue = {2},
	pages = {529--552},
	numpages = {0},
	year = {2013},
	publisher = {American Physical Society},
	doi = {10.1103/RevModPhys.85.529},

}

@ARTICLE{Heslot1994,
	author = {F. Heslot and T. Baumberger and B. Perrin and B. Caroli and C. Caroli},
    title = {Creep, stick-slip, and dry-friction dynamics: Experiments and a heuristic model},
	year = 1994,
	journal = {Phys. Rev. E},
	volume = 49,
	pages = {4973-4988},
    doi={10.1103/PhysRevE.49.4973}
}

@article{Baumberger1999,
Title = {Physical analysis of the state- and rate-dependent friction law. II. Dynamic friction},
author = {Baumberger, T. and Berthoud, P. and Caroli, C.},
journal = {Phys. Rev. B},
volume = {60},
issue = {6},
pages = {3928--3939},
numpages = {0},
year = {1999},
publisher = {American Physical Society},
doi = {10.1103/PhysRevB.60.3928},

}

@article{Baumberger2006,
	title={Solid friction from stick--slip down to pinning and aging},
	author={Baumberger, Tristan and Caroli, Christiane},
	journal={Advances in Physics},
	volume={55},
	number={3-4},
	pages={279--348},
	year={2006},
	publisher={Taylor \& Francis},
    doi = {10.1080/00018730600732186}
}

@Article{Marone1998,
	author = {Marone, C.},
	title = {Laboratory-derived friction laws and their application to seismic faulting},
	journal =  {Ann. Revs. Earth \& Plan. Sci.}, 
	volume = {26}, 
	pages = {643-696},  
	year = {1998},
    doi={10.1146/annurev.earth.26.1.643}
}

@ARTICLE{Scholz1998,
	author = {Scholz, Christopher H.},
	year = {1998},
	journal = {Nature},
	volume = {391},
	pages = {37},
	title = {Earthquakes and Friction Laws},
    doi={10.1038/34097}
}

@article{Rice1983,
 author={ Rice, J. R.  and Ruina,, A. L.},
 journal =    {J. Appl. Mech.},
 title = {Stability of Steady Frictional Slipping},
 volume = {50}, 
 pages = {343-349}, 
 year ={1983},
 doi={https://doi.org/10.1115/1.3167042}
}

@article{Desplanques2015,
 ISSN = {19463979, 19463987},
 author = {Yannick Desplanques},
 journal = {SAE International Journal of Materials and Manufacturing},
 number = {1},
 pages = {98--103},
 publisher = {SAE International},
 title = {Amontons-Coulomb Friction Laws, A Review of the Original Manuscript},
 urldate = {2024-06-17},
 volume = {8},
 year = {2015},
 doi={10.4271/2014-01-2489}
}

@article{Hutchings2016,
title = {Leonardo da Vinci's studies of friction},
journal = {Wear},
volume = {360-361},
pages = {51-66},
year = {2016},
issn = {0043-1648},
doi = {https://doi.org/10.1016/j.wear.2016.04.019},
author = {Ian M. Hutchings}
}

@article{Plati2022,
  title = {Collective Drifts in Vibrated Granular Packings: The Interplay of Friction and Structure},
  author = {Plati, A. and Puglisi, A.},
  journal = {Phys. Rev. Lett.},
  volume = {128},
  issue = {20},
  pages = {208001},
  numpages = {6},
  year = {2022},
  publisher = {American Physical Society},
  doi = {10.1103/PhysRevLett.128.208001},
}

@article{Marchand2020,
  title = {Roughness-Induced Friction on Liquid Foams},
  author = {Marchand, Manon and Restagno, Fr\'ed\'eric and Rio, Emmanuelle and Boulogne, François},
  journal = {Phys. Rev. Lett.},
  volume = {124},
  issue = {11},
  pages = {118003},
  numpages = {6},
  year = {2020},
  publisher = {American Physical Society},
  doi = {10.1103/PhysRevLett.124.118003},
}

@article{Petrillo2020,
author={Petrillo, Giuseppe
and Lippiello, Eugenio
and Landes, Fran{\c{c}}ois P.
and Rosso, Alberto},
title={The influence of the brittle-ductile transition zone on aftershock and foreshock occurrence},
journal={Nature Communications},
year={2020},
day={15},
volume={11},
number={1},
pages={3010},
issn={2041-1723},
doi={10.1038/s41467-020-16811-7},
}

@article{Yan2023,
author={Yan, Caishan
and Chen, Hsuan-Yi
and Lai, Pik-Yin
and Tong, Penger},
title={Statistical laws of stick-slip friction at mesoscale},
journal={Nature Communications},
year={2023},
day={05},
volume={14},
number={1},
pages={6221},
issn={2041-1723},
doi={10.1038/s41467-023-41850-1},
url={https://doi.org/10.1038/s41467-023-41850-1}
}

@book{gere2008mechanics,
  title={Mechanics of Materials, SI Edition},
  author={Gere, J.M. and Goodno, B.J.},
  isbn={9780495438076},
  lccn={2008923451},
  url={https://books.google.fr/books?id=O7Xy0JaT5qUC},
  year={2008},
  publisher={Cengage Learning}
}

@phdthesis{fulleringer2014contribution,
  title={Contribution to the study of friction phenomena: application to paper materials},
  author={Fulleringer, Nicolas},
  year={2014},
  school={Universit{\'e} de Grenoble}
}

@article{garoff2004friction,
  title={Friction hysteresis of paper},
  author={Garoff, Niklas and Fellers, Christer and Nilvebrant, Nils-Olof},
  journal={Wear},
  volume={256},
  number={1-2},
  pages={190--196},
  year={2004},
  publisher={Elsevier}
}

@inproceedings{de1997determination,
  title={Determination of the friction of paper and board},
  author={De Silveira, G and Hutchings, IM},
  booktitle={The fundamentals of papermaking materials. Transactions of the XIth fundamental research symposium, Cambridge. FRC, Manchester},
  pages={1329--1353},
  year={1997}
}

@book{borch2001handbook,
  title={Handbook of Physical Testing of Paper: Volume 2},
  author={Borch, Jens and Lyne, M Bruce and Mark, Richard E and Habeger, Charles},
  year={2001},
  publisher={Crc Press}
}

@article{kawashima2008paper,
  title={Paper friction at the various measuring conditions-effect of relative humidity},
  author={Kawashima, Noriaki and Sato, Jun and Yamauchi, Tatsuo},
  journal={Sen'i Gakkaishi},
  volume={64},
  number={11},
  pages={336--339},
  year={2008},
  publisher={The Society of Fiber Science and Technology, Japan}
}

\clearpage
\appendix

\begin{figure}[h]
\includegraphics[width=0.99\columnwidth]{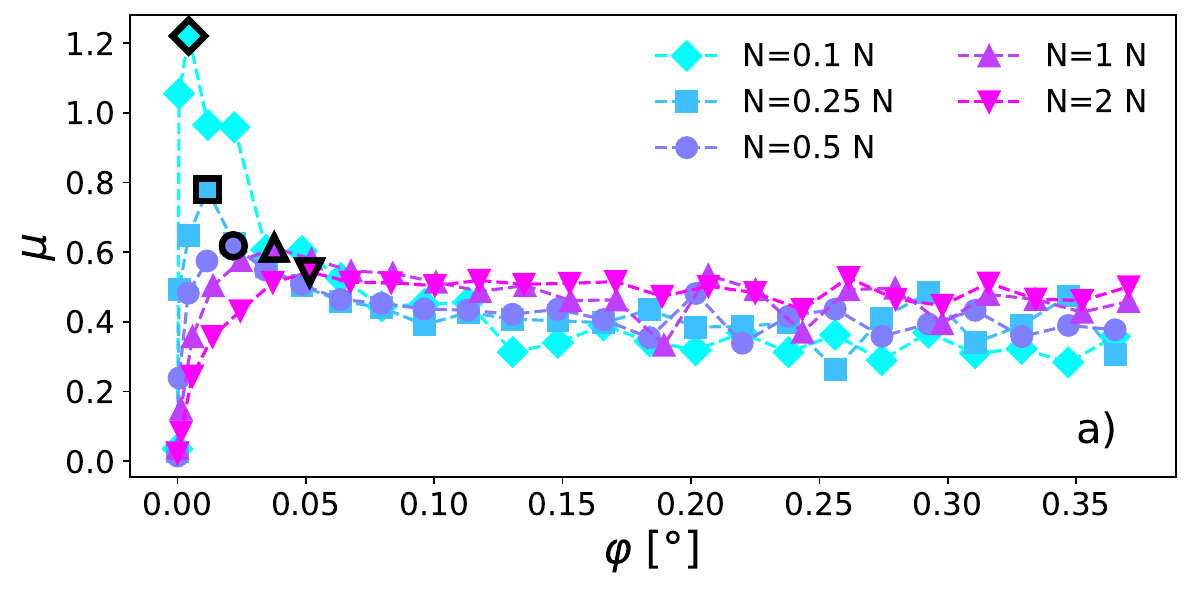}
\includegraphics[width=0.99\columnwidth]{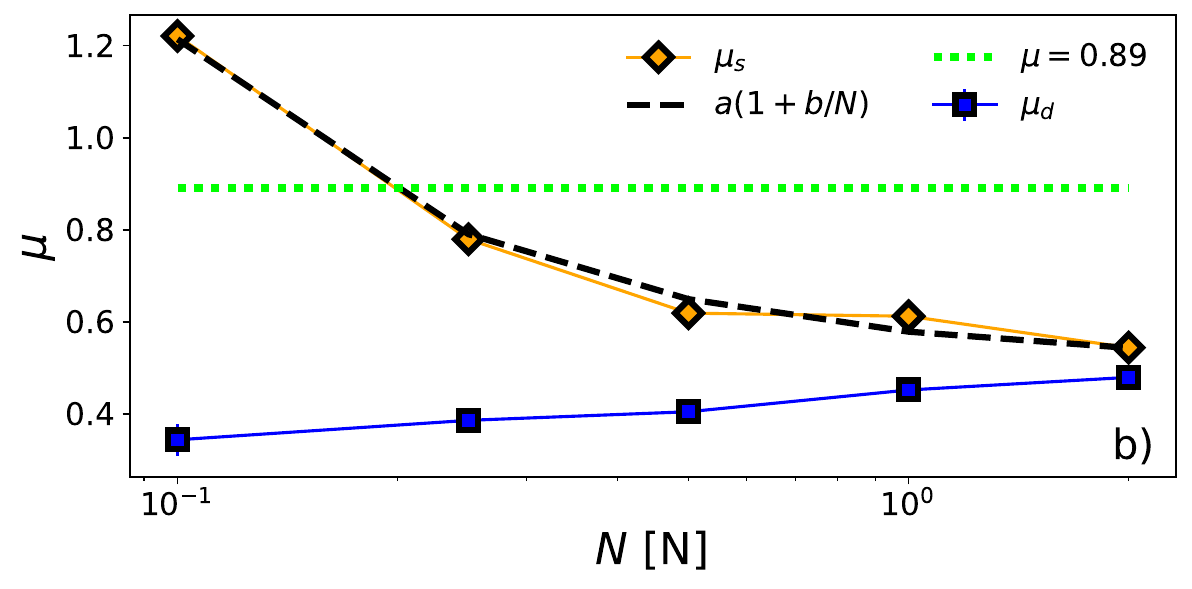}
\includegraphics[width=0.99\columnwidth]{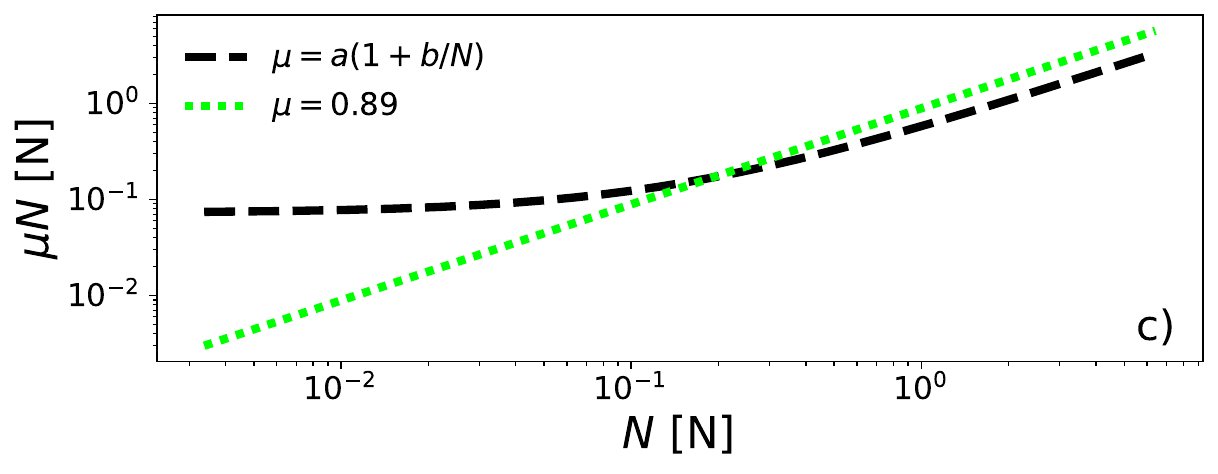}
\caption{a) Instantaneous friction coefficient as a function of the traveled angle for different imposed normal forces. The static friction coefficient is estimated from the local maxima of the curves, which are highlighted with black bordered markers. b) Static and dynamic friction coefficients as a function of the imposed normal force. The behavior of $\mu_s$ is well described by Eq. \eqref{eq::crass}.  \modif{The friction coefficient $\mu=0.89$ obtained by fitting $F_\text{max}(d)$ via Eq.~\eqref{eq::model} (Fig.~\ref{fig:MaxVsD}) falls within the range of the measured $\mu_s(N)$.} The dynamic friction coefficient has been obtained by averaging the curves $\mu(\varphi)$ over the range $\varphi\in[0.3,120]^{\circ}$. Data have been obtained by making two Post-it paper surfaces slide on each other in a plane-plane geometry of a rheometer. These measurements have been performed with an imposed angular velocity $\Omega=0.1$ rad/s. \modif{c)  Static friction threshold forces as a function of $N$ for the full range of normal forces expected inside the interleaved assembly. We compare here the predictions based on the constant friction coefficient $\mu=0.89$, and those based on a normal-force-dependent friction coefficient following Eq.~\eqref{eq::crass}. This analysis suggests that the effectiveness of the constant friction coefficient is due to the balance between the overestimation of tangential forces at large $N$ and the underestimation at small $N$.}} \label{fig:appendFrott}
\end{figure}

\section{Single sheets characterization}\label{app:singlesh}

During our experimental investigation we characterized the mechanical properties of our Post-it\textregistered~sheets.

\textit{Friction coefficients}---First, we measured the paper-on-paper friction coefficients as a function of the normal force between two Post-it surfaces. This was done through a rheometer (Anton Paar, MCR 302) by covering the two sides of a plane-plane geometry with paper coming from the Post-it\textregistered~sheets (adherence was assured by an adhesive tape). This geometry involves a circular contact area under a normal load $N$ and a shear stress originated by the rheometer that imposes a rotation at angular velocity $\Omega$. To convert the measured torque $\tau$ into a friction coefficient taking into account the shear imposed by relative rotation (and not by translation as in usual tribometers), we used the following formula
\begin{equation}
    \mu=\frac{3}{2}\frac{\tau}{RN}
\end{equation}
where $R=25$ mm is the radius of the circular contact area. The above formula takes into account that the measured torque can be expressed as $\tau=\int_0^R dr\int_0^{2\pi}d\theta\mu N/(\pi R^2)r^2$, where the integrand represents the contribution of an infinitesimal contact area element at distance $r$ from the center to the total torque.
Our measurements were done starting at rest and recording the measured torque as a function of the traveled angle $\varphi$. From the resulting $\mu(\varphi)$ curves, we extract the static friction coefficient $\mu_s$ from the initial peak and the dynamic one $\mu_d$ averaging $\mu(\varphi)$ over the range $\varphi\in[0.3,120]^{\circ}$ (see Fig.~\ref{fig:appendFrott}a). By repeating these measurements for different imposed normal forces $N\in[0.1,2]$ N, we obtained $\mu_s(N)$ and $\mu_d(N)$ shown in Fig.~\ref{fig:appendFrott}b. We observe that $\mu_s$ decreases as a function of $N$ while $\mu_d$ remains relatively constant showing only a slight increase over the whole range of $N$. The behavior of $\mu_s(N)$ can be explained based on the previous work by
Crassous et al. \cite{crassous1999humidity} which, based on experimental evidences, proposed the following dependence of the static friction coefficient on the normal force:
\begin{equation}\label{eq::crass}
\mu(N)=a(1+b/N).    
\end{equation}
The divergence of the friction coefficient for vanishing normal load is due to adhesion, which determines a non-zero friction force even when $N\to 
0$. 
As shown in Fig.~\ref{fig:appendFrott}b, the measured paper-on-paper static friction coefficient shows a good agreement with Eq.~\eqref{eq::crass} having $a=0.507$ and $b=0.143$.
As for explaining the slight increase of $\mu_d$ as a function of $N$, further investigations, which fall outside  the scope of this paper, would be needed. Possible scenarios include a crossover between elastic and plastic contacts or the formation of wear particles~\cite{achanta2010scale}.
 \modif{The horizontal dotted line in Fig.~\ref{fig:appendFrott}b represents the friction coefficient $\mu=0.89$ obtained by fitting $F_\text{max}(d)$ via Eq.~\eqref{eq::model} (Fig.~\ref{fig:MaxVsD}). We observe that it falls within the range of the measured $\mu_s(N)$. A better understanding of why a constant friction coefficient can be representative for our interleaved system can be obtained by considering the predicted static friction threshold forces $\mu N$ on individual sheets inside the assembly. These thresholds depend on the normal force $N$, which spans a wide range of values inside the assembly, and also on the separation distance $d$. A lower bound for this quantity is $T^*/\mu$, which represents the friction force due to adhesion between two pages of the assembly~\cite{taub2021nonlinear}. An upper bound can be estimated from Eq.~\eqref{eq::normal} by considering the innermost sheet of the assembly $N(n=1)$ at the onset of detachment, $d\sim13.5$ mm. This results in a range of normal forces spanning more than three decades, $N\in[0.0034,6.4]$ N. Comparing the predictions for $\mu N$ obtained with $\mu=0.89$ and those obtained using Eq.~\eqref{eq::crass} over this range (Fig.~\ref{fig:appendFrott}c), we observe that a constant friction coefficient overestimates the static thresholds at large $N$ while underestimating them at small $N$. This suggests that the effectiveness of the fitted constant $\mu$ can be interpreted as resulting from a balance between these two opposing mismatches.  }

\begin{figure}
\includegraphics[width=0.99\columnwidth]{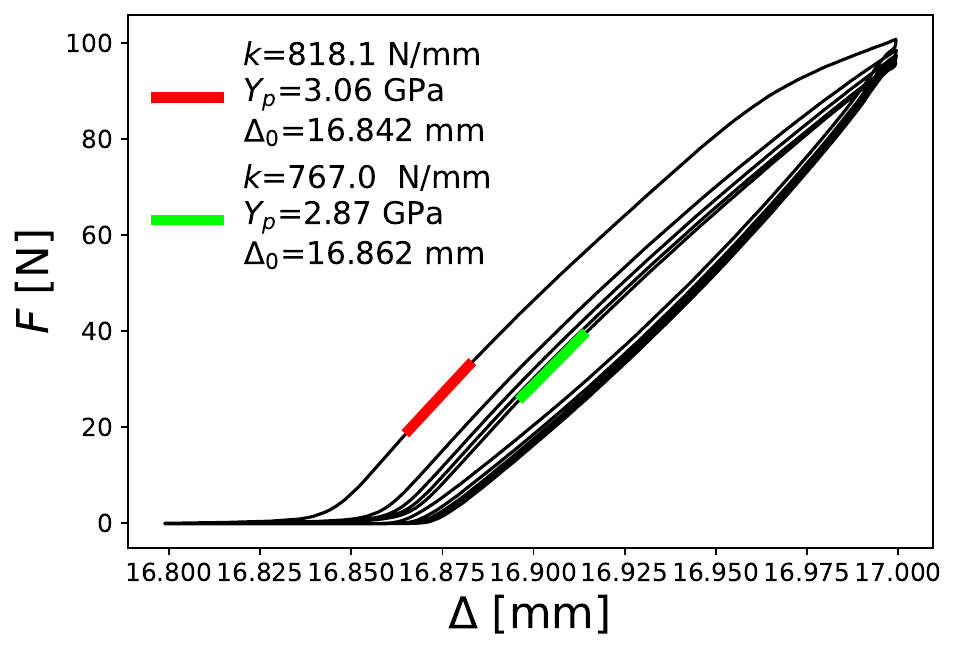}
    \caption{Measured traction force as a function of the separation distance between the two clamping points for a single Post-it paper sheet. The vertical displacement is imposed at $V=1$ mm/min and consists of 5 back-and-forth cycles.  Different portions of the detachment branches can be fitted as $F=k\Delta+F_0$. Then, one can take $Y_p=k\Delta_0/\epsilon W$, where $\Delta_0=k/F_0$. } \label{fig:appendYoung}
\end{figure}

\emph{Single sheet under tensile stress}---We also estimated the Young's modulus of the paper by doing traction tests on single paper sheets. We imposed a vertical displacement back and forth to a $\epsilon=0.1$ mm thick, $L=16.8$ mm long, and $W=45$ mm wide sheet of Post-it\textregistered~paper.  In Fig.~\ref{fig:appendYoung} we show an illustrative example of this analysis where the traction force $F$ is plotted as a function of the distance $\Delta$ between the clamping points. The vertical displacement is imposed at $V=1$ mm/min and consists of 5 back-and-forth cycles. The Young's modulus $Y_p$ can be estimated by fitting different portions of the detachment branches (i.e. the red and green lines in the figure) as $F=k\Delta+F_0$. Then, one can take $Y_p=k\Delta_0/\epsilon W$, where $\Delta_0=k/F_0$. By repeating this analysis for different sheets we obtained values of $Y_p\in[2,3]$ GPa regardless of the imposed $V\le 1$ mm/min.

\section{Image processing}\label{app:improc}

\begin{figure}
\includegraphics[width=0.99\columnwidth]{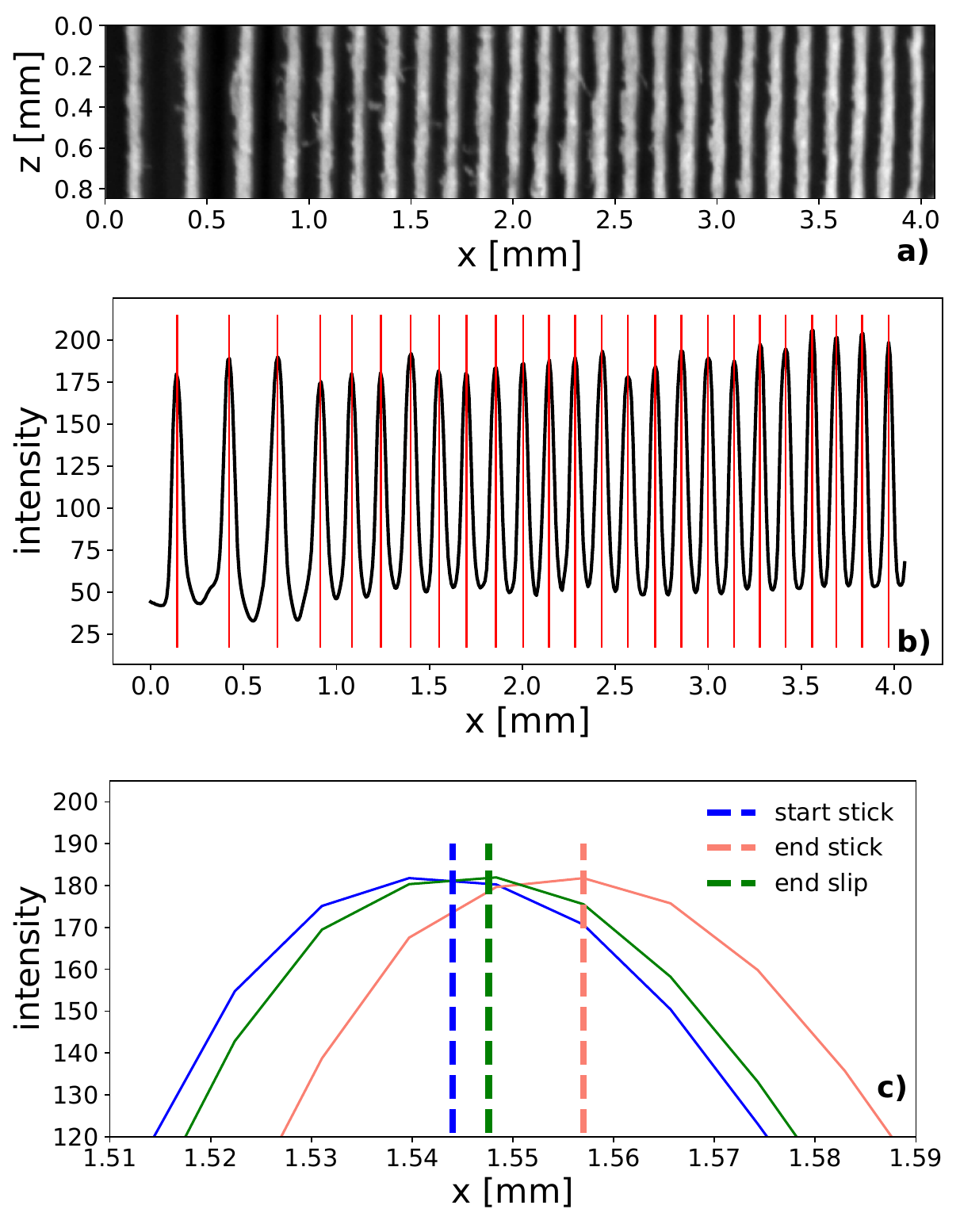}
    \caption{a) Illustrative picture of the subregion of the images analyzed in Sec.~\ref{sec:video}.  b) Intensity averaged over $z$ as a function of the horizontal pixel coordinate for one frame at the beginning of the stick phase. The detected horizontal sheet positions are highlighted by red vertical  lines. c) Zoom on pixel intensities for a single sheet taken from three different frames: one at the beginning of the stick, one at the end of the stick, one at the end of the slip. The detected horizontal sheet positions are highlighted by dashed vertical lines. The position clearly shifts inward during the stick and come back close to the initial one after the slip.} \label{fig:appendImProc}
\end{figure}

The analysis carried on in Sec.~\ref{sec:video} is based on the detection of the instantaneous horizontal positions $x(n,t)$ of the sheets. These are extracted from the subregion of our frames corresponding to the white mark in Fig.~\ref{fig:ImageProc}a and shown in Fig.~\ref{fig:appendImProc}a.  \modif{In order to identify the horizontal position of the sheets we first detect
the local maxima of the pixel intensity averaged over $z$ in the aforementioned subregion. Then, we consider a 9-pixel-wide interval centered around these peaks and fit the light intensity in this region with a parabola. The sheet positions are then identified by the horizontal coordinate the parabola's vertices. The main advantage of this method with respect to a simple peak detection is that it exploits average information from a whole pixel interval, meaning that the position of the sheets can be defined without being constrained to discrete pixel coordinates. This is particularly useful for reconstructing the horizontal motion of the sheets during the very short stick phases occurring in the late state of the detachment process.}
 In Fig.~\ref{fig:appendImProc}c, we show that our method allows following the sheet horizontal displacement during stick-slip events. 

\begin{figure}
\includegraphics[width=0.6\columnwidth]{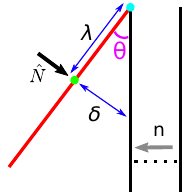}
    \caption{An outer sheet (drawn in red) is schematized as a cantilever beam with the upper end fixed at the interleaving point (cyan circle). The application point of $\hat{N}$ is represented by the green circle and is located at a distance $\lambda$  from the interleaving point. We consider $\delta$ to be the typical distance that the force application point needs to displace in order for an outer sheet to adhere to the inner ones (drawn in black).} \label{fig:bendSheet}
\end{figure}

\section{Sheet bending}\label{app:bend}

As discussed in Sec.~\ref{sec:effnom} and sketched in Fig.~\ref{fig:Setup}b,  the portion of the sheet joining the clamping point to the start of the interleaved part makes an angle $\theta(n)$ with the vertical axis. Without any additional confining force in the horizontal direction, a sheet would not adhere to the rest of the assembly, making an angle $\theta(n)$ with $z$ not only at the clamping point but also at the interleaving point. This condition is indeed realized by the outer sheet in the assembly that has only the horizontal component of gravity as a confinement (see Fig.~\ref{fig:Setup}a). More internal sheets bend  due to the normal force exerted by the outer sheets, thus resulting more parallel to the vertical axis. In order to estimate a typical bending force, we will consider sheets as cantilever beams with the upper end fixed at the interleaving point (see Fig.~\ref{fig:bendSheet}). Through beam mechanics~\cite{gere2008mechanics} we obtain that to displace the application point of the force by a distance $\delta$ we need $\hat{N}=3\delta IY_p/\lambda^3$, where $I=\epsilon^3W/12$ is the second moment of inertia of the sheet and $\lambda$ is the distance from the interleaved point and the application point of $\hat{N}$. 
Then, the 
displacement needed to make the sheet $n$
adhere to the inner sheets of the assembly is $\delta=\lambda\tan{\theta(n,d)}=\lambda n\epsilon/d$. The application point $\lambda$ of the normal force is not easy to estimate, we expect it to be not too distant from the interleaving point and having the interleaved length $L-d$ as an upper bound. Putting together the above expressions we can check that,   
taking outer sheets $n\in [15,25]$, the interleaving point within $d\in [13,18]$ mm and the Young's modulus in the range $Y_p\in[2,3.8]$ GPa, we find that to a fitted $\hat{N}=0.57$ N corresponds 
a $\lambda=\sqrt{nWY_p\epsilon^4/4d\hat{N}}\in[1.8,3.8]$ mm. If we consider the whole interval of fitted values $\hat{N}\in [0.48,0.77]$ N we obtain $\lambda\in[1.56,4.14]$ mm. These correspond to reasonable ranges of values for $\lambda$, which tells us that the estimated threshold normal forces $\hat{N}$ are compatible with the typical force needed to make a sheet adhere to the inner ones.

\section{\modif{Variation of fitted parameters}}\label{app:fitPar}

\begin{figure}
\includegraphics[width=0.9\columnwidth]{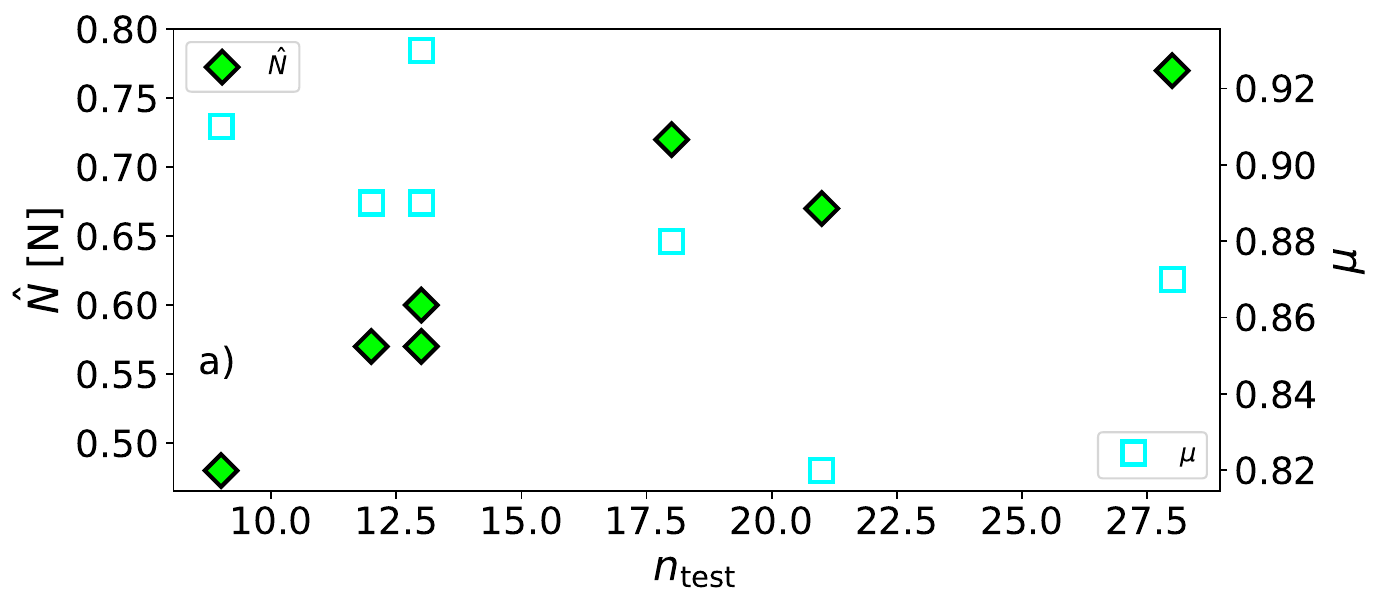}
\includegraphics[width=0.9\columnwidth]{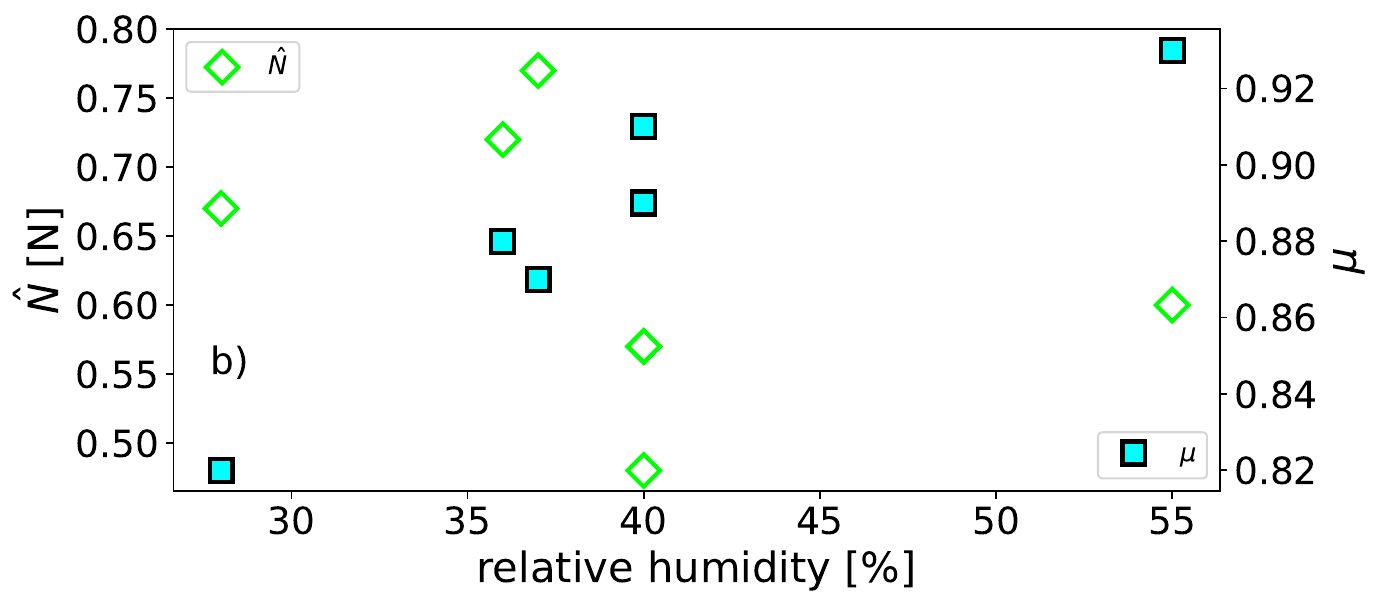}
\includegraphics[width=0.45\columnwidth]{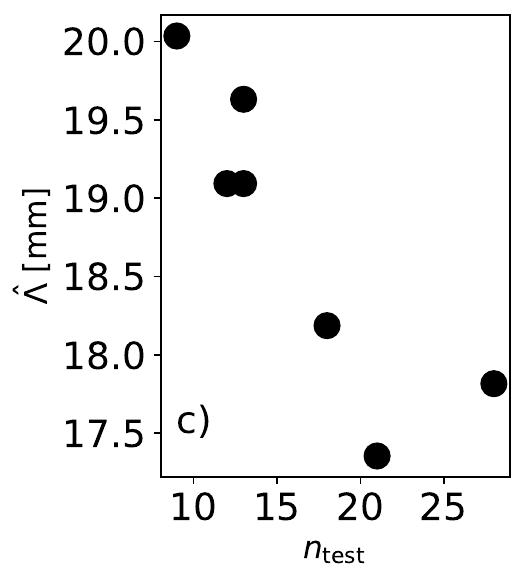}
\includegraphics[width=0.45\columnwidth]{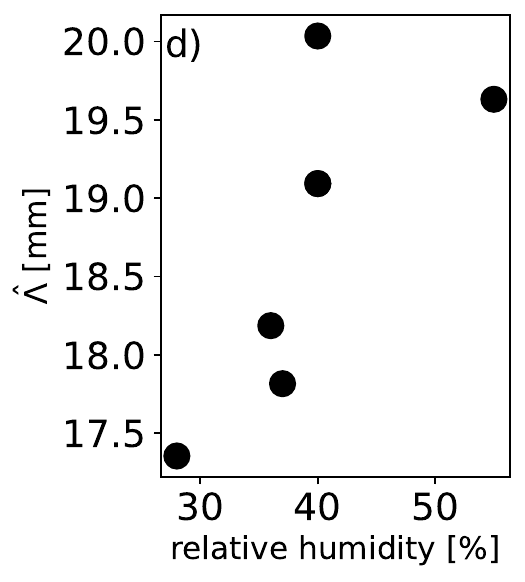}
    \caption{\modif{a) Fitted $\hat{N}$ and $\mu$ from the analysis of Sec.~\ref{sec:effnom} as a function of the number $n_\mathrm{test}$ of traction tests in the range $d\in[13,18]$ mm performed before data acquisition. b) Same data plotted as a function of the measured ambient relative humidity. Characteristic distance $\hat{\Lambda}$ as a function of $n_\mathrm{test}$ (panel c) and ambient relative  humidity (panel d). The behavior of $\hat{\Lambda}$ indicates that aging tends to soften the interleaved assemblies, while humidity makes them stiffer.}} \label{fig:fitParam}
\end{figure}

\modif{Figures~\ref{fig:fitParam}a and b complement the analysis of Sec.~\ref{sec:effnom} by showing the behavior of the fitted $\mu$ and $\hat{N}$ as a function of the measured ambient relative humidity and the number of previous traction tests $n_\mathrm{test}$ performed before each experiment. This last variable provides a simple way to quantify the degree of aging of the samples. We find that $\hat{N}$ and $\mu$ exhibit clear increasing trends with $n_\mathrm{test}$ and humidity, respectively. The increase of $\hat{N}$ with $n_\mathrm{test}$ can be explained by considering that repeated tests result in looser assemblies, so that individual sheets require a higher horizontal compression to contribute to the collective stiffness. The increase of $\mu$ with humidity recovers the standard behavior expected for paper-on-paper friction~\cite{borch2001handbook,kawashima2008paper}. At the same time, in a much less pronounced fashion, $\mu$ appears to show a slight decreasing trend when plotted against $n_\mathrm{test}$, as does $\hat{N}$ when plotted against relative humidity. The weak trend of $\mu$ in Fig.~\ref{fig:fitParam}a indicates that the friction coefficient is mainly affected by aging during the first runs and remains relatively stable after the smallest considered value $n_\mathrm{test}=9$. The weak trend of $\hat{N}$ in Fig.~\ref{fig:fitParam}b simply shows that humidity does not substantially affect the interplay between horizontal confinement and effective stiffness. Finally, in Figs.~\ref{fig:fitParam}c and d, we show the behavior of the characteristic length $\hat{\Lambda}=2\mu \epsilon M^2/\ln(1+2\mu\hat{N}/T^*)$ as a function of $n_\mathrm{test}$ and relative humidity. This characteristic length defines the dimensionless variable $\hat{d}=d_v/\hat{\Lambda}$ used for the data collapse in Fig.~\ref{fig:finalScaling}. A larger $\hat{\Lambda}$ therefore reduces $\hat{d}$ and in turn increases the effective stiffness of the system. The observed behavior shows that aging acts on $\hat{N}$ and $\mu$ in a way that decreases $\hat{\Lambda}$ (system's softening), while humidity has the opposite effect.}

\clearpage

\begin{table}[htbp]
    \centering
    \caption{List of symbols}
    \label{tab:symbols}
    \begin{tabular}{>{\centering\arraybackslash}m{3cm} l}
        \toprule
        \textbf{Symbol} & \textbf{Description} \\
        \midrule
        $n$             & Sheet index, from center ($n=1$) to outer part ($n=M$) \\
        $M$             & Number of sheets per block (half the total, $2M=50$) \\
        $\epsilon$      & Sheet thickness (0.1 mm) \\
        $W$             & Sheet width (45 mm) \\
        $L$             & Distance from clamping point to sheet end (57 mm) \\
        $L_t$           & Total length of a block (75 mm) \\
        $d$             & Separation distance from clamping to interleaving point \\
        $d_0$           & Initial separation distance at start of experiment (13 mm) \\
        $\theta(n)$     & Angle between vertical axis and sheet $n$ at clamping point \\
        $V$             & Imposed detachment velocity \\
        $F$             & Total measured traction force \\
        $\mathcal{T}$   & Total frictional force opposing detachment (model prediction) \\
        $\mathcal{T}_0$ & Force offset due to zero-setting of force sensor at $d=d_0$ \\
        $T^*$           & Friction force between the two pages held together by adhesion force \\
        $N(n)$          & Normal compression force felt by sheet $n$ from outer sheets \\
        $\mathcal{N}$   & Average normal force acting on the assembly \\
        $\hat{N}$       & Normal force threshold separating stiff from soft sheets \\
        $\mu$           & Static friction coefficient between sheet surfaces (mechanical model)\\
        $\mu_s$         & Static friction coefficient (single sheet characterization, spring-block model)\\
        $\mu_d$         & Dynamic friction coefficient (single sheet characterization, spring-block model)\\
        $\alpha$        & Dimensionless amplification parameter, $\alpha = 2\epsilon\mu M^2/d$ \\
       $K$        & Effective stiffness of the assembly \\
        $K_p$           & Stiffness of a single paper sheet under tensile stress \\
        $K_{\rm eff}/2$   & Effective stiffness per contributing sheet couple \\
        $m$             & Block mass (spring-block model) \\
        $w$             & Normal load on block (spring-block model) \\
        $k$             & Spring stiffness (spring-block model)\\
        $u(t)$          & Spring elongation (spring-block model) \\
        $u_0$           & Spring elongation at end of slip (spring-block model) \\
        $f(t)$          & Elastic force in the spring, $f=ku$ (spring-block model) \\
        $v$        & Velocity imposed at the free-end of the spring (spring-block model) \\ 
        $F_p$, $F_{\max}$ & Local maximum (peak) force within a stick-slip event \\
        $F_v$           & Local minimum (valley) force within a stick-slip event \\
        $\Delta F$      & Stick-slip amplitude, $\Delta F = F_p - F_v$ \\
         $A$             & Prefactor linking stick-slip amplitude to normal force, $\Delta F = A\mathcal{N}$ \\
        $d_p$           & Separation distance at force peak within a stick-slip event \\
        $d_v$           & Separation distance at force valley (start of stick phase) \\
        $t_v$           & Time at start of stick phase \\
        $t_{\rm st}$    & Duration of stick phase \\
        $t_{\rm sl}$    & Duration of slip phase \\
        $\Lambda_{\rm st}$ & Stick length (displacement during stick phase) \\
        $\Lambda_{\rm sl}$ & Slip length (displacement during slip event) \\

        $x(n,t)$        & Instantaneous horizontal position of sheet $n$ \\
        $\hat{x}(n)$    & Mean horizontal position of sheet $n$ over stick phase \\
        $\delta x(n,t)$ & Horizontal displacement of sheet $n$ relative to its mean \\
        $\sigma_x(n)$   & Mean absolute horizontal displacement during stick phase \\
        $M_{\rm eff}$   & Effective number of sheets contributing to global stiffness \\
        $\hat{d}$       & Adimensional separation, $\hat{d}=\ln(1+2\mu\hat{N}/T^*)/\alpha$ \\
        $\hat{\Lambda}$ & Characteristic length, $\hat{\Lambda}=2\mu\epsilon M^2/\ln(1+2\mu\hat{N}/T^*)$ \\
        $Y_p$           & Young's modulus of the paper \\
        $Y_{\rm eff}$   & Effective Young's modulus used in stiffness model \\
        $\tau$          & Torque measured by rheometer \\
        $R$             & Radius of circular contact area in rheometer (25 mm) \\
        $\Omega$        & Angular velocity imposed by rheometer \\
        $\varphi$       & Traveled angle in rheometer measurement \\
        $a,\,b$         & Fitting parameters for $\mu_s(N)=a(1+b/N)$ \\
        $I$             & Second moment of inertia of a sheet, $I=\epsilon^3 W/12$ \\
        $\lambda$       & Distance from interleaving point to force application point \\
        $\delta$        & Displacement to make a sheet adhere to inner ones \\
        \bottomrule
    \end{tabular}
\end{table}

\end{document}